\documentclass[aip,pop,reprint,floatfix,footinbib]{revtex4-1}
\usepackage{graphicx}
\usepackage[breaklinks=true,colorlinks=true,linkcolor=blue,urlcolor=blue,citecolor=blue]{hyperref}

\newcommand{\vc}{\mathbf}

\begin{document}
\title{Theory of the Electron Sheath and Presheath} 

\author{Brett Scheiner}
\email[]{brett-scheiner@uiowa.edu}
\affiliation{Department of Physics and Astronomy, University of Iowa, Iowa City, Iowa 52242, USA}

\author{Scott D. Baalrud}
\affiliation{Department of Physics and Astronomy, University of Iowa, Iowa City, Iowa 52242, USA}

\author{Benjamin T. Yee}
\affiliation{Sandia National Laboratories, Albuquerque, New Mexico 87185, USA}

\author{Matthew M. Hopkins}
\affiliation{Sandia National Laboratories, Albuquerque, New Mexico 87185, USA}

\author{Edward V. Barnat}
\affiliation{Sandia National Laboratories, Albuquerque, New Mexico 87185, USA}

\date{\today}

\begin{abstract}
Electron sheaths are commonly found near Langmuir probes collecting the electron saturation current. The common assumption is that the probe collects the random flux of electrons incident on the sheath, which tacitly implies that there is no electron presheath and that the flux collected is due to a velocity space truncation of the electron velocity distribution function (EVDF). This work provides a dedicated theory of electron sheaths, which suggests that they are not so simple. Motivated by EVDFs observed in Particle-In-Cell (PIC) simulations, a 1D model for the electron sheath and presheath is developed. In the model, under low temperature plasma conditions ($T_e\gg T_i$), an electron pressure gradient accelerates electrons in the presheath to a flow velocity that exceeds the electron thermal speed at the sheath edge. This pressure gradient generates large flow velocities compared to what would be generated by ballistic motion in response to the electric field. It is found that in many situations, under common plasma conditions, the electron presheath extends much further into the plasma than an analogous ion presheath. PIC simulations reveal that the ion density in the electron presheath is determined by a flow around the electron sheath and that this flow is due to 2D aspects of the sheath geometry. Simulations also indicate the presence of ion acoustic waves excited by the differential flow between electrons and ions in the presheath which result in sheath edge fluctuations. The 1D model and time averaged PIC simulations are compared and it is shown that the model provides a good description of the electron sheath and presheath. 
\end{abstract}

\maketitle

\section{INTRODUCTION}
Sheaths, which are present at essentially any plasma boundary, are one of the most fundamental structures in plasma physics and have been studied extensively\cite{2013PPCF...55i3001R}.  Sheaths play the important role of maintaining global current balance, allowing the existence of a quasineutral plasma. At floating boundaries the sheath is ion rich (an ion sheath), providing a thin positive space charge layer that limits the electron losses to the boundary. Not all sheaths need to be ion rich. Sheaths near small electrodes, such as those around Langmuir probes, can be electron rich (electron sheaths) when the electrode is biased positive with respect to the plasma potential. Due to the requirements of global current balance, electron sheaths are possible only near electrodes that are small enough that the ratio of their area to the plasma chamber wall area satisfies $A_E/A_\textrm{w}<\sqrt{2.3 m_e/m_i}$ where $A_E$ and $A_w$ are the effective surface areas for collecting charged particles at the electrode and the wall, respectively \cite{2007PhPl...14d2109B}. The effect of the electrode-to-wall area ratio on the sheath form has been experimentally verified \cite{2014PhPl...21j3512B}. 

Electron sheaths are most commonly encountered around Langmuir probes collecting the electron saturation current\cite{1926PhRv...28..727M,1961JAP....32.2512M, 1962JAP....33.3094M}, but are also encountered around plasma contactors\cite{1992PhFlB...4.3847A,1991JPhD...24.1789S}, tethered space probes\cite{1996JGR...10117229S}, and in laser accelerated plasmas\cite{vosoughian2015enhancement}. Electron sheaths have also been observed to play a role in probe induced particle circulation in dusty plasmas crystals\cite{1998PhRvL..80.4189L} and are also important for providing electrons with the energy needed to ionize neutral atoms in the formation of anode spots\cite{2009PSST...18c5002B, 2008PSST...17c5006S}. The present understanding of electron sheaths from Langmuir probe theory is comprised of the following: 1) The electron sheath collects the random flux of electrode-directed electrons\cite{2005PhPl...12e5502H,1926PhRv...28..727M}. This flux is given by $\Gamma_R=\frac{1}{4}n_eA_{E}\bar{v_e}$, where $\bar{v}_e=\sqrt{8T_e/\pi m_e}$ is the mean electron velocity, $T_e$ is the electron temperature in eV, and $m_e$ is the electron mass. 2) Since the flux collected is random, the EVDF is a half Maxwellian at the electron sheath edge\cite{1961JAP....32.2512M, 1962JAP....33.3094M}. 3) The electron sheath analog of the Bohm criterion is trivially satisfied\cite{1991JPhD...24..493R,2006PSST...15..773C} because the truncation of the EVDF at the sheath edge provides the required flow moment. Presheaths have not been considered. 4) Ions near the electron sheath follow a Boltzmann density profile $n_i=n_o\exp(-e\phi/T_i)$, where $n_i$ is the ion density, $n_o$ is a reference density at $\phi = 0$, $T_i$ is the ion temperature in eV, and $\phi$ is the electrostatic potential\cite{2008PhPl...15g3507S}. In this paper we consider a dedicated theory of the electron sheath and find that each of these assumptions need to be revisited.

In a recently submitted paper Yee et al.~\cite{2015arXiv150805971Y} (hereafter YE), it was shown that under low temperature plasma conditions the electron sheath is accompanied by a presheath where an electron flow of approximately an electron thermal speed is generated due to pressure gradients. This presheath was shown to extend well into the bulk plasma, even extending beyond the range of an analogous ion presheath. In YE, results from particle-in-cell simulations with direct simulation Monte-Carlo collisions (PIC-DSMC) showed that the EVDF near the sheath edge was a flowing Maxwellian. In the present paper, we present a new theoretical model for the electron sheath and presheath based on observations from these recent experiments and simulations. This new theory shows that the electron fluid flow exceeds the electron thermal speed by the sheath edge, satisfying an electron sheath analog of the Bohm criterion. The 1D model describes the electron flow as being pressure gradient driven, this is significantly different than the electric field driven flow in ion presheaths. This presheath pressure gradient generates large flow velocities over regions with little change in potential. The effect on the bulk plasma can be significant even with small gradients in the plasma potential. 

Although the 1D model provides an accurate characterization of many aspects of the electron sheath and presheath, some aspects of the 2D simulations are not captured by the 1D theory. The need to satisfy global current balance usually results in electron sheaths occurring only around small electrodes, hence the infinite planar picture common to 1D models is not perfect. In the simulation the electron presheath causes the ion flow to be redirected around the small electrode, resulting in a significantly different situation than that described by a Boltzmann density profile. Analysis of the PIC simulations reveal that the ion density is only accurately described when the ion flow is taken into account.  

The previous simulations in YE showed that the electron sheath edge exhibits fluctuations on the order of 1 MHz. Two-dimensional FFTs of the ion density show these fluctuations are ion acoustic waves excited by the differential flow between fast electrons, and ions in the electron presheath. The time dependence of the ion density fluctuations closely correlate with the sheath edge fluctuations. The sheath edge position fluctuations may explain the current fluctuations previously observed for probes biased above the plasma potential\cite{2014PhPl...21j3512B,1981PlPh...23..325G,1978JPSJ...44..991D}. In addition, these fluctuations may contribute to an effective electron-ion collision rate in the electron presheath through instability enhanced collisions\cite{2010PhPl...17e5704B}.

This paper is organized as follows. Section II discusses the implications of different EVDF models on the electron sheath and presheath, and develops a fluid-based approach motivated by PIC simulations. Section IIIA describes the PIC simulations and results. Section IIIB provides a comparison between simulations and the model, Sec. IIIC focuses on the ion behavior in the presheath, and Sec. IIID on the time-dependent aspects as well as instabilities. Concluding statements are made in Sec. IV.

\section{MODEL}

\subsection{Conventional kinetic models}

The present understanding that a Langmuir probe collects the random thermal flux in electron saturation is based on the assumption that the electron sheath interfaces with the bulk plasma without a presheath. A direct consequence is that the EVDF under this model is a Maxwellian that has no flow shift, but is truncated at zero velocity at the sheath edge; see Fig.~1. In this section, the consequences of the conventional assumption are first explored. This demonstrates that even under the conventional assumption a finite electric field, and hence a presheath should be expected. Since a presheath generates flow, a flowing truncated Maxwellian such as shown in Fig.~1 might be suggested as an appropriate model. However, recent PIC simulations have shown that the expected flow speed should be very fast, approaching the electron thermal speed by the sheath edge. Furthermore, the observed distribution was a flowing Maxwellian; see Fig.~2. Motivated by these simulation results, a fluid-based model is developed in Sec. IIB. This provides a model for the minimum flow speed to be expected at the sheath edge, a model for pressure driven flow in the presheath, and a model for electric field driven flow in the sheath. 

\begin{figure}
\centering
\includegraphics[scale=.4]{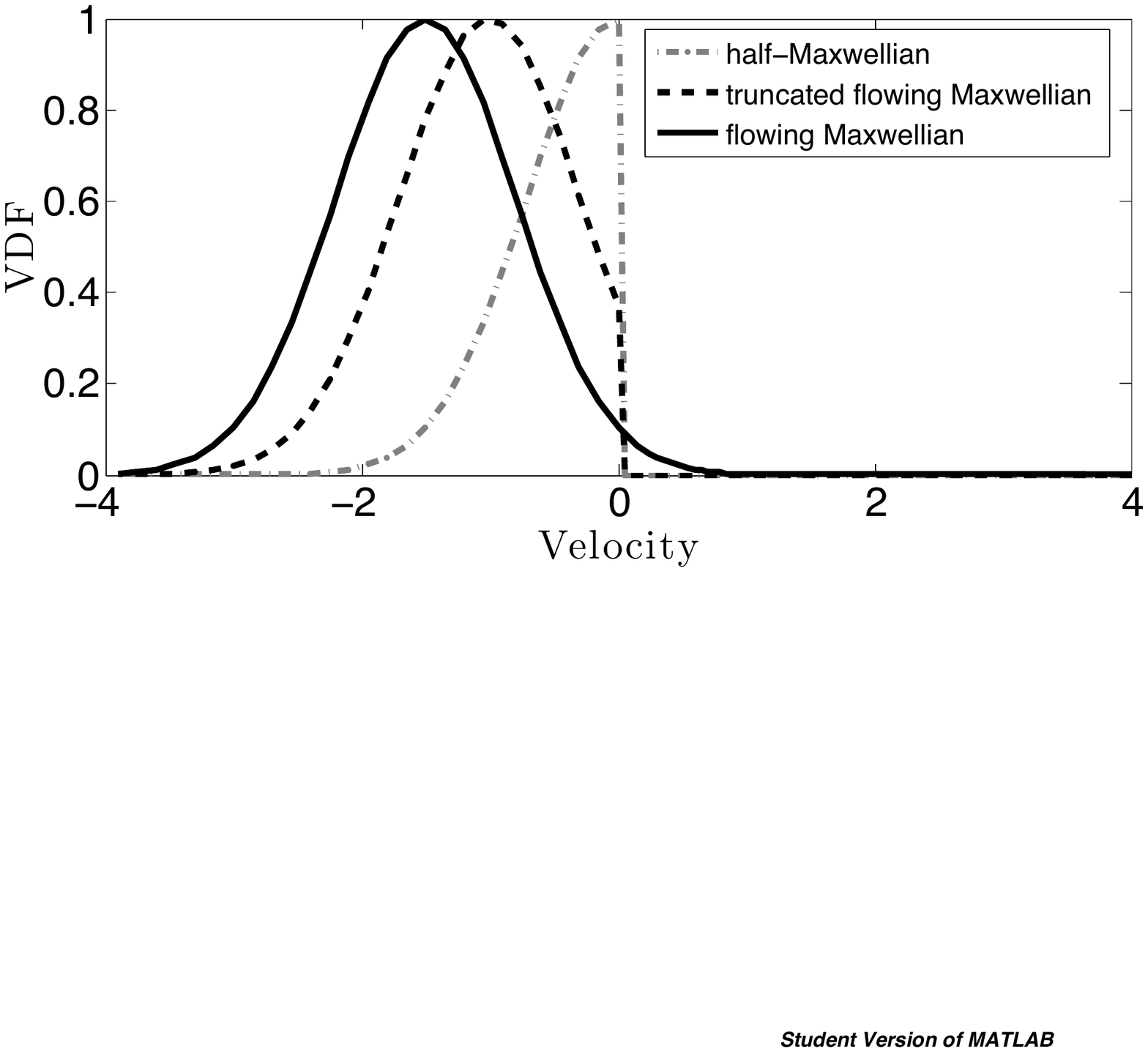}
\caption{Three different model velocity distribution functions. The half Maxwellian corresponding to a collisionless electron-rich sheath with no presheath, the truncated flowing Maxwellian corresponding to a collisionless sheath with a presheath, and the flowing Maxwellian corresponding to the collisional sheath with presheath.  }
\end{figure}

\begin{figure}
\centering
\includegraphics[scale=.45]{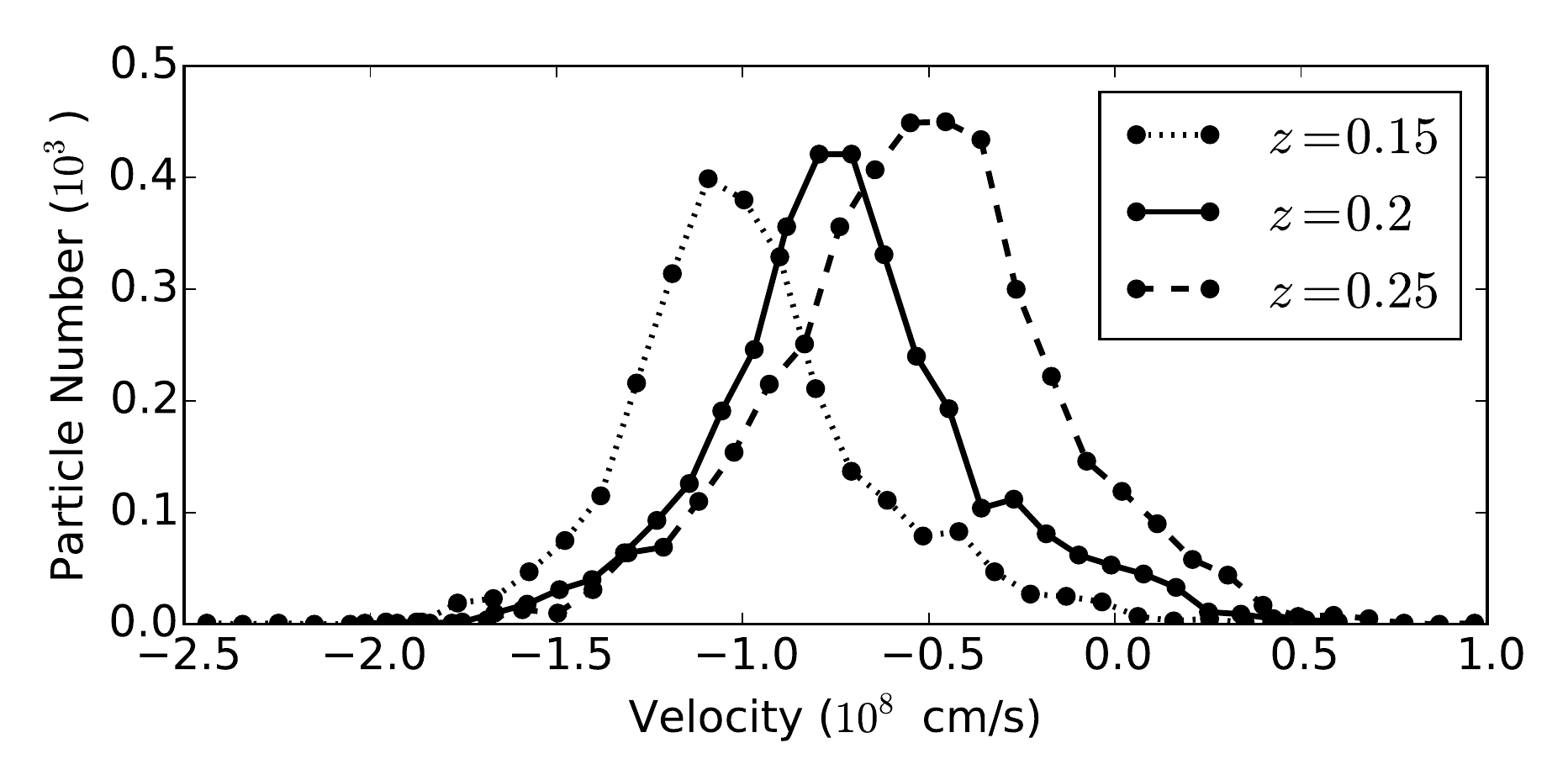}
\caption{ PIC simulation results showing that the EVDF can be modeled as a flowing maxwellian in the electron sheath and presheath. The sheath edge is at $y\approx0.25$ cm.}
\end{figure}

Consider, briefly, an implication of the conventionally assumed picture where an electron sheath interfaces directly with the bulk plasma without a presheath. In this case the sheath edge EVDF is a truncated Maxwellian, however the assumed EVDF in the bulk plasma is Maxwellian. This picture does not allow the matching of the bulk plasma to the sheath since such a transition would break flux conservation. Since the bulk plasma EVDF is typically close to maxwellian, stress gradients and friction terms in the momentum equation are negligible, so any change in electron flux must be due to an electric field. This difficulty of transitioning between the bulk plasma and sheath suggests a need for a plasma region where an electric field allows the the acceleration of electrons providing an electron flux at the sheath edge. Two implications of this directly hint at the need to revisit the half-Maxwellian assumption: 1) A non-zero electric field at the sheath edge implies that the electrons have a flow due to the electric force exerted on them, and 2) a finite electric field also implies that there is a density gradient in the quasineutral presheath that can give rise to pressure gradient induced flows, e.g., for Boltzmann ions $n_i=n_o \exp(-e\phi/T_i)$, $dn_i/dz=en_iE/T_i$.  The following section will show that the latter effect is the dominant mechanism for electron-flow generation in the presheath when $T_e \gg T_i$.

\subsection{New fluid model}

Motivated by PIC simulation results showing that not only is there a flow shift, but that the EVDF is Maxwellian at the sheath edge, this section describes a fluid-based model of the electron sheath and presheath. The acceleration mechanism is found to be a pressure gradient. The implications of this pressure driven flow are that the electrons can achieve large flow velocities even over regions where the potential varies by a small amount.  In this section, it is shown that the electrons accelerated by this presheath must enter the sheath with a flow speed exceeding the electron thermal speed, a result that may be considered an electron sheath analog of the Bohm criterion. We will refer to this as the electron sheath Bohm criterion (the speed that must be satisfied will be referred to as the electron sheath Bohm speed, denoted $v_{eB}$). A 1D model is developed for the density, flow, and potential profiles in the presheath and sheath regions. 

\subsubsection{Sheath edge}
For the purposes of modeling the presheath and sheath edge, consider a model that describes electrons with continuity and momentum equations, assuming that the plasma is generated at a rate proportional to the density. Ions are assumed to obey a Boltzmann density, $n_i=n_{o}\exp(-e\phi/T_i)$, where $n_{o}$ is the density in the bulk plasma. These equations are supplemented with Poisson's equation and an isothermal closure for electrons. Since we are concerned with the presheath, the quasineutrality condition applies, and the density gradient can be written as $dn_e/dy=en_iE/T_i$. Inserting this into the momentum equation \begin{equation}
 V_e\frac{d V_e}{d y}=-\frac{e}{m_e}E-\frac{T_e}{m_e n_e}\frac{d n_e}{d y}-V_e(\nu_R+\nu_s)
 \end{equation}
 shows that the pressure gradient term is $T_e/T_i$ times larger than the electric field term. Here, in Eq. (1), $V_e$ denotes the first moment of the EVDF, and $\nu_R$ and $\nu_s$ denote the collision frequencies due to momentum transfer collisions and particle source rate respectively.  In typical low temperature plasmas $T_e/T_i \sim 10-100$, hence the flow is dominantly pressure driven. This situation makes a significant contrast with ion sheaths, where instead the ion pressure gradient term is $T_i/T_e \ll 1$ smaller than the electric field term.

This model can be used to determine the conditions on the electron flow velocity at the sheath edge. Expanding the charge density about a position at the sheath edge $\rho(\phi)=\rho(\phi_0) +d\rho/d\phi|_{\phi =\phi_0}(\phi-\phi_0)+...$, and defining the sheath edge as the location where neutrality breaks down, gives a common definition of the sheath edge\cite{1991JPhD...24..493R}  $\big|d\rho/d\phi|_{\phi =\phi_0}\big|>0$. This requirement, which is known as the sheath criterion, can be rewritten as $\sum_s q_s dn_s/dy\le0$ where the sum is over each plasma species. The Bohm criterion for a fluid model can be obtained by inserting the fluid equations into this form of the sheath criterion\cite{2011PSST...20b5013B,2015PPCF...57d4003B}. For the electron sheath, consider a thin region near the sheath edge where the source and collision terms can be neglected. The electron continuity equation, along with Eq.(1) and the Boltzmann density relation for ions, then imply the following electron sheath analog of the Bohm criterion 

\begin{equation}
V_e\ge\sqrt{\frac{T_e+T_i}{m_e}}\equiv v_{eB}.
\end{equation} 
A similar electron sheath Bohm criterion was previously found\cite{2012PhPl...19h3507L}, but was not derived from consideration on the EVDF. The electron sheath Bohm speed in Eq.(2) is approximately $\sqrt{m_i/m_e}$ greater than the ion sound speed, which is the ion flow generated in an ion presheath. Because this is significantly faster than the ion sound speed, the differential flow between ions and electrons is expected to excite ion acoustic instabilities in the electron presheath. This will be studied in Sec. IIID.  Next we will consider analytic solutions for the plasma parameter profiles in the presheath and sheath. 

\subsubsection{Presheath}
In this subsection the properties of the quasineutral presheath are explored. A mobility limited flow equation is derived for the electron fluid. The equations for velocity and potential profiles are solved in a region in the vicinity of the sheath edge and analytic solutions are found for the cases of constant mean free path and constant collision frequency. The solutions demonstrate that large flow velocities are obtained over regions in which there is a small potential gradient. From these solutions it is found that in some cases the electron presheath has an extent that is $\sqrt{m_i/m_e}$ longer than that of the analogous ion presheath, and under more typical low temperature plasma conditions the presheath is $\sim6$ times longer than the ion presheath. This means that the electron sheath can perturbed the bulk plasma over a few centimeters under typical laboratory conditions.   

Starting with the quasineutrality condition on the density gradient, and using the first two fluid moment equations, an electron mobility limited flow equation is obtained,
\begin{equation}
  V_e=-\mu_e\bigg(1-\frac{V_e^2}{v_{eB}^2}\bigg)E.
 \end{equation}
This equation is analogous to the ion mobility limited flow equation, but where $\mu_e=e(1+T_e/T_i)/[m_e(\nu_R+2\nu_s)]$ is the electron mobility. When compared with the the ion mobility in an ion presheath with a common collision frequency due to volume ionization of neutrals, the electron mobility greatly exceeds ion mobility $\mu_e\approx \frac{T_e m_i} {T_im_e}\mu_i$. 

Next, consider a region in the vicinity of the sheath edge that is thin enough that an assumption of constant flux, $n_eV_e=n_ov_{eB}$, is accurate.  Here $n_o$ is the density at the sheath edge. Using this form of the electron density along with the Boltzmann density for ions in Poisson's equation gives
\begin{equation}
\bigg(\frac{\lambda_{D_e}^2}{l^2}\bigg)\frac{d^2(e\phi/T_e)}{d(y/l)^2}=-\bigg(e^{-e\phi/T_i}-\frac{v_{eB}}{V_e}\bigg),
\end{equation}
where $l$ is the presheath length scale. Taking the quasineutral limit $\lambda_{D_e}/l\to0$ gives the potential as a function of flow velocity
\begin{equation}
\phi=-\frac{T_i}{e}\ln\bigg(\frac{v_{eB}}{V_e}\bigg).
\end{equation}
This form of the potential along with the mobility limited flow in Eq. (3) results in a differential equation for the flow velocity in terms of spatial position, 
\begin{equation}
\frac{dy}{dV_e}=\frac{v_{eB}^2-V_e^2}{(\nu_R+\nu_s) V_e^2}.
\end{equation}
The solution to this differential equation along with  Eq. (5) gives the flow and potential profile. This differential equation has an ion sheath analog\cite{2005ppdm.book.....L}, which has analytic solutions\cite{2010PhDT.......242B} for 1) the case of constant mean free path, $\nu=V_e/l$, and 2) constant collision frequency, $\nu=v_B/l$. For the case of constant mean free path the flow velocity is 

\begin{equation}
\frac{V_e}{v_{eB}}=\exp\bigg\{\frac{1}{2}-\frac{y}{l}+\frac{1}{2}W_{-1}\bigg[-\exp\bigg(2\frac{y}{l}-1\bigg)\bigg]\bigg\}
\end{equation}
where $W_{-1}$ is the $-1$ branch of the Lambert W function\cite{Corless96onthe}. Eq. (5) gives the potential profile

\begin{equation}
-\frac{e\phi}{T_i}=\frac{y}{l}-\frac{1}{2}-\frac{1}{2}W_{-1}\bigg[-\exp\bigg(2\frac{y}{l}-1\bigg)\bigg].
\end{equation}

For the constant $\nu$ case the flow and potential profiles are 
\begin{equation}
\frac{V_e}{v_{eB}}=1-\frac{y}{2l}\Bigg(1+\sqrt{1-\frac{4l}{y}}\Bigg)
\end{equation} 
and

\begin{equation}
-\frac{e\phi}{T_i}=\textrm{arccosh}\bigg(1-\frac{y}{2l}\bigg).
\end{equation}

The flow velocity and potential profiles for these two cases are shown in Fig.~3. These show that large flows are obtained over regions with shallow potential gradients and little change in potential. Flow velocities of this magnitude are not seen in the ion presheath. 

\begin{figure}[h!]
\centering
\includegraphics[scale=0.45]{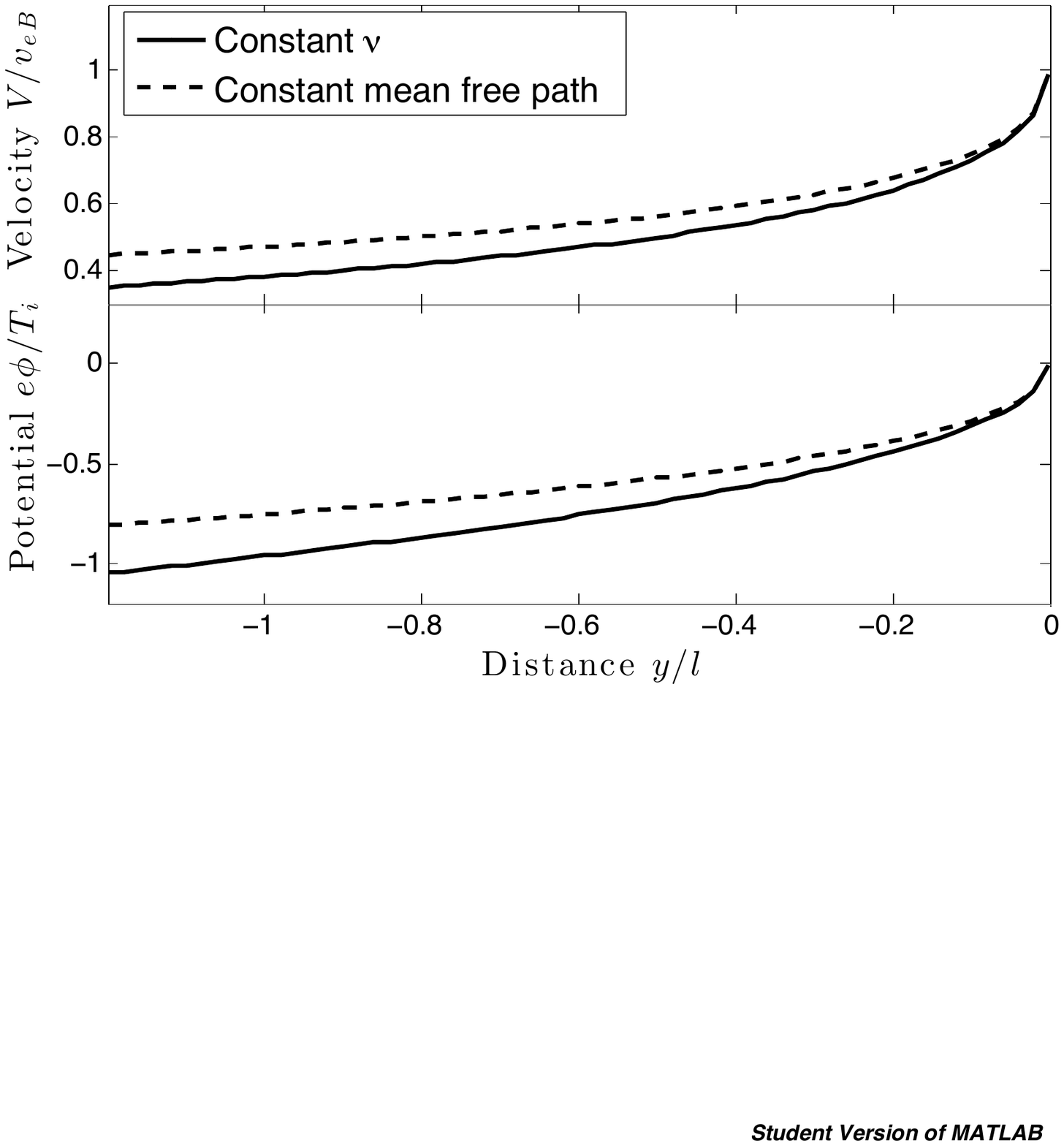}
\caption{The flow velocity (top) and potential profile (bottom) for the presheath for the cases of constant collision frequency and constant mean free path from Eqs. (7)-(10).}
\end{figure}

For the case of constant collision frequency the characteristic length scale of electron and ion presheaths can be compared explicitly. Two cases are considered; A) a plasma where volume ionization is the dominant effect, and B) a helium plasma with momentum transfer collisions and no volume ionization. 

A) Consider a plasma where the dominant collision process is volume ionization so that $\nu_s$ is the same for ions and electrons. If the sheath attached to the electrode is an ion sheath the presheath length scale would be $l_i=c_s/\nu$, while if the sheath were an electron sheath the presheath length scale is $l_e=v_{eB}/\nu$. The ratio of these two length scales is 
\begin{equation}
\frac{l_e}{l_i}=\frac{v_{eB}}{c_s}=\sqrt{\frac{T_e+T_i}{m_e}\frac{m_i}{T_e}}\approx\sqrt{\frac{m_i}{m_e}}.
\end{equation} 
This suggests that the characteristic length scale of an electron presheath can be more than an order of magnitude longer than an ion presheath. A typical ion sheath length scale in low temperature plasma experiments is $\sim1$ cm \ \cite{2002PhRvL..89n5001O}, which means for the case of an argon plasma, where $\sqrt{m_i/m_e}\approx270$, the implied presheath length scale would be $l_e\approx270$ cm. This is longer than the scale of many plasma experiments, so it would be expected that the presheath would fill approximately half the experiment length\cite{2005PSST...14..201O}. 

B) In this case the collision frequencies are different, using $\nu_s=n_g K_{s}$ the ratio of presheath length scales is 
\begin{equation}
\frac{l_e}{l_i}=\frac{v_{eB}}{c_s}\frac{\nu_i}{\nu_e}\approx\sqrt{\frac{m_i}{m_e}}\frac{K_i}{K_e}
\end{equation}
where $K_s$ is the rate constant for collisions between neutral helium and species s. When the temperature is small the rate constant is 
\begin{equation}
K_s(U)\approx U\sigma_s(U).
\end{equation}
Using this approximation the rate constants were estimated using flow speeds representative of typical presheath velocities, $U=v_{eB}/2$ for the electron presheath and $U=c_s/2$ for the ion presheath. For the calculation of $K_e$ the total momentum cross section for $\text{e}^- +\text{He}$ collisions was obtained from LXcat\cite{lxcat}, while for the calculation of $K_i$ the cross section for $\text{He}^+ + \text{He}$ elastic and charge exchange collisions were considered\cite{1957JChPh..26.1272C}. For the $\text{He}^+ +\text{He}$ cross sections the values at 4 eV were extrapolated to 0 eV as has been previously done\cite{2015PSST...24d4005S}, this was due to a lack of data within this range of energies. The cross sections used are shown in Fig.~4. The ratio of presheath length scales shown in Fig.~5 suggest that, for this case, the electron presheath is approximately six times longer than the ion presheath. These values are in good agreement with the estimated presheath lengths determined from density measurements in YE where the electron and ion presheaths were measured to be approximately 25mm and 6mm respectively.

\begin{figure}
\includegraphics[scale=.44]{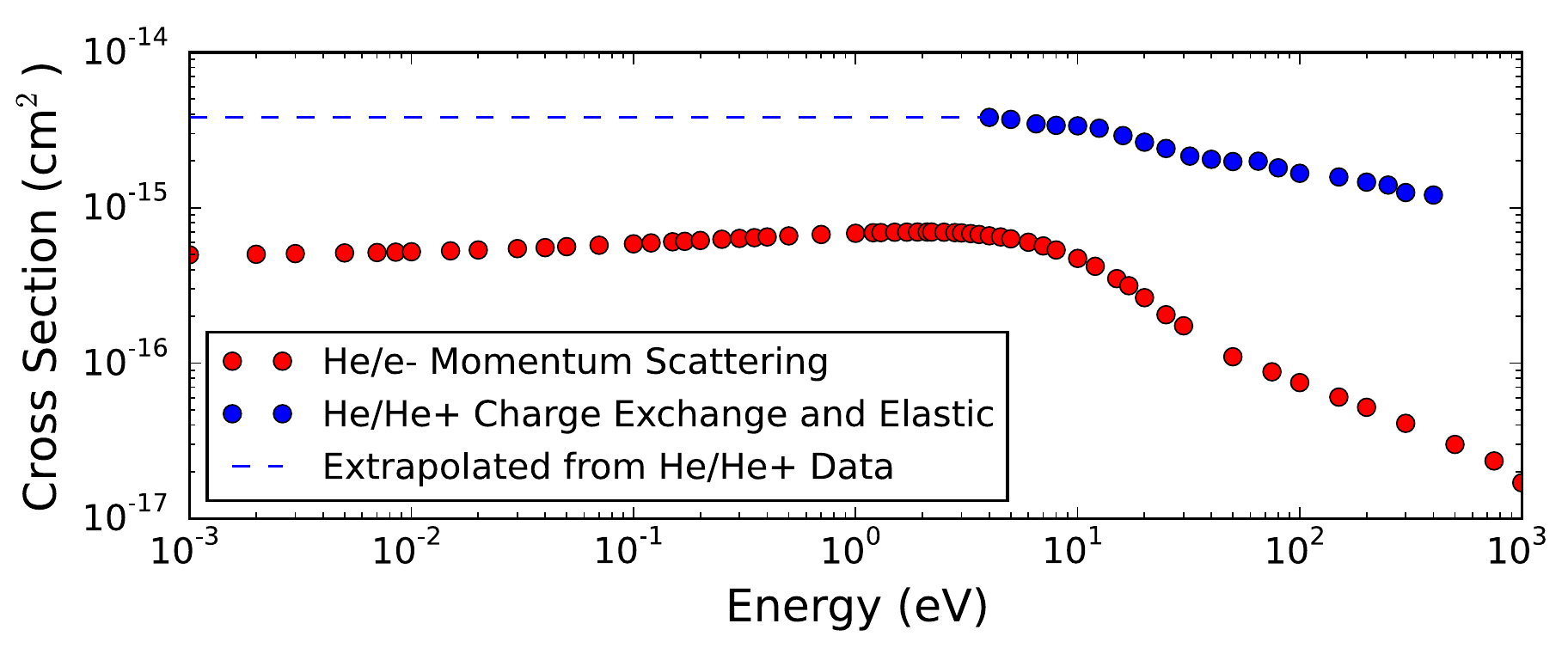}
\caption{The neutral helium-electron momentum scattering, and neutral helium-helium ion charge exchange and elastic cross sections used for the calculation of the rate constant. Note the extrapolation of the helium cross section, this was necessary due to lack of data at low energies. }
\end{figure}

\begin{figure}
\includegraphics[scale=.44]{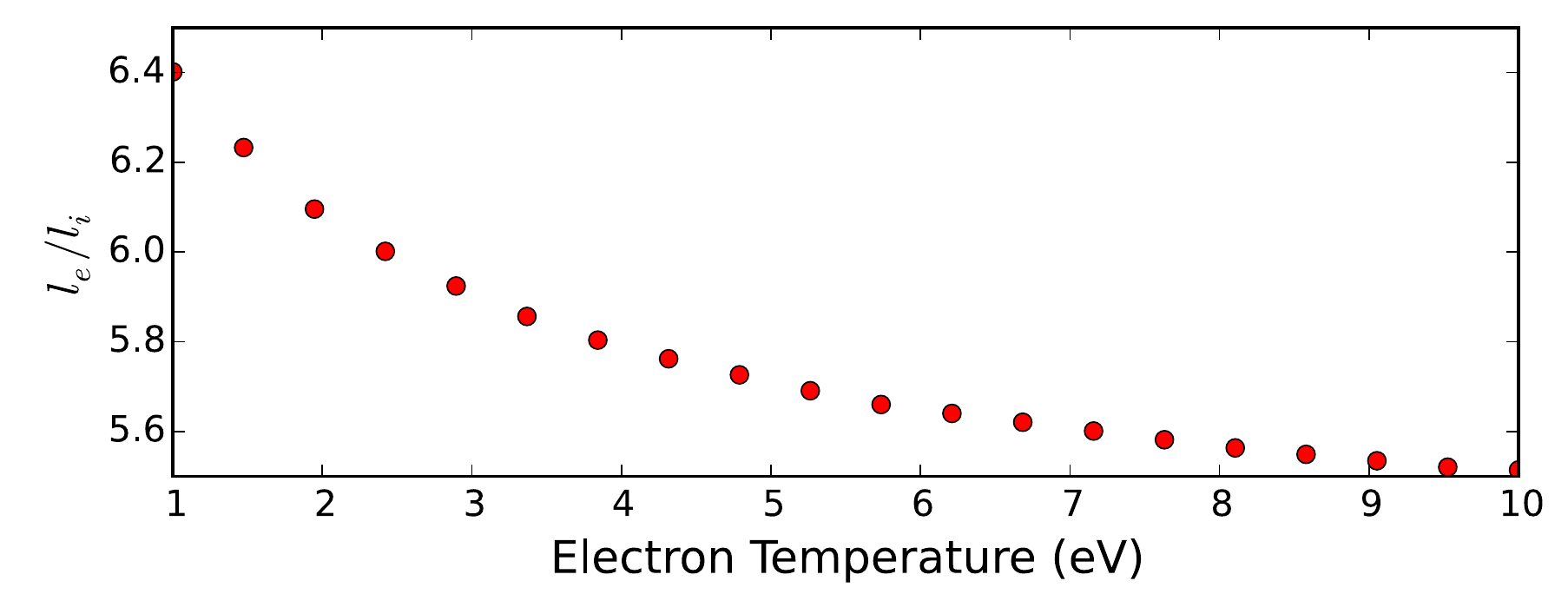}
\caption{The ratio of presheath length scales as a function of $T_e$.}
\end{figure}
These results differs substantially from the conventional picture of electron sheaths, which are thought to be local phenomena. Instead, this suggests that electron sheaths often influence a plasma globally. This suggestion will be considered further in Sec. IIIB with PIC simulations where there is no presheath volume generation of plasma and hence $\nu_s=0$.

\subsubsection{Sheath} 

 For the electron sheath, the sheath-presheath transition is a region where the flow switches from being pressure driven to electric field driven. In the thin sheath region, the collision and source terms can be neglected \footnote{The physical justification for dropping the collision term is that the sheath is so thin that electrons will flow to the boundary without suffering a collision, that is the sheath scale is much smaller than the collisional mean free path}. This provides a way to determine a relation between the flow velocity and potential, which shows that at small potentials the pressure represents a significant correction to the electron ballistic motion, while at high potentials the 3/4 power law scaling of the Child-Langmuir law \cite{1911PhRvI..32..492C} is recovered.  

Under the assumptions mentioned above, combining the continuity and momentum equations, integrating and matching the sheath edge conditions results in 
\begin{equation}
\bigg(\frac{V_e}{v_{eB}}\bigg)^2-2\ln\bigg(\frac{V_e}{v_{eB}}\bigg)=\frac{2e\phi}{T_e}+1.
\end{equation}
The second term on the left hand side is the contribution due to the electron pressure. The solution to this equation can be written in terms of the Lambert W function, however the logarithmic electron pressure term is at most a correction of $\approx 20 \%$ in the sheath and drops off at higher flow velocity. In the asymptotic limit this term is negligible (see Appendix A). 

Using the asymptotic solution for $V_e$, enforcing that the electron density within the sheath obeys flux conservation  ($n_e(\phi)V_e(\phi)=n_ov_{eB}$), and neglecting the ion density, which decreases exponentially with increasing potential, Poisson's equation can be written as 
\begin{equation}
\frac{d^2\phi}{dy^2}=\frac{4\pi e n_o}{\sqrt{1+\frac{2e\phi}{T_e}}}.
\end{equation}
Integrating twice with respect to y gives 
\begin{equation}
\frac{y_{flowing}}{\lambda_{D_e}}=0.79\bigg(\frac{e\Delta\phi}{T_e}\bigg)^{3/4}
\end{equation}
which is the same as what is obtained for the ion sheath\cite{2005PhPl...12e5502H}. A different relation for the electron sheath has been previously given\cite{2005PhPl...12e5502H}\footnote{The numerical factor in \cite{2005PhPl...12e5502H} is $0.32/\sqrt{\alpha}$ where $\alpha$ is a correction factor, greater than unity, due to a dip in front of the sheath reducing the density at the sheath edge. This dip is not observed in the simulations in the following section and the value of $\alpha=1$ is used here.}, here the sheath scaling was given as 

\begin{equation}
\frac{y_{truncated}}{\lambda_{D_e}}=0.32\bigg(\frac{e\Delta\phi}{T_e}\bigg)^{3/4}.
\end{equation}

 This different numerical factor is due to the random flux assumption. Comparing Eq. (16) and (17) gives the correction to the sheath scale 
\begin{equation}
\frac{y_{flowing}}{y_{truncated}}=2.47,
\end{equation}
which suggests that the electron sheath is more than twice as thick as previously thought. In Sec. IIIB this relation is found to be in excellent agreement with simulations.

\section{SIMULATIONS}
The model in the previous section assumed a 1D planar electron sheath. In this section the model is tested using 2D PIC simulations. These show that the electron sheath has some inherently 2D features that are not accurately captured by the model. In particular, the ion density is found to be determined by a 2D ion flow velocity profile around the electron sheath. Nevertheless, basic features of the 1D model, such as the minimum electron flow speed at the sheath edge, are found to accurately represent the simulations. Modifications of the 1D theory to address 2D ion flow are found to lead to improvements in the predicted presheath profiles. The PIC simulations also exhibit fluctuations in sheath thickness. In Sec. IIID, evidence is shown that these fluctuations are ion acoustic waves excited in the presheath. 
\subsection{Aleph}
The simulations were performed using the PIC-DSMC code Aleph. Aleph is an electrostatic PIC code that utilizes direct simulation Monte Carlo (DSMC) kinetic techniques\cite{bird1998molecular} for interparticle collisions. The algorithm represents a plasma by evolving electrostatically coupled computational particles in time and computes the particle positions and velocities on an unstructured mesh in 1D, 2D, or 3D, each with three velocity components\cite{2012CoPP...52..295T}. In our simulations a 2D triangular mesh with a resolution of approximately $0.7 \lambda_{De}$ was used. The simulations utilized a 7.5 cm $\times$5 cm domain with one reflecting and three grounded absorbing boundaries. An electrode of length 0.25 cm was embedded  in an absorbing wall adjacent to the reflecting boundary and was separated from the wall by a gap filled with a dielectric of length 0.2 cm, see Fig.~10 for an image of the simulation domain. The domain, which was set up to resemble experiments on a reduced scale, was filled with a helium plasma that was continuously generated in a source region $\sim$4 cm above the electrode, and expanded to fill the domain. Plasma in this region was sourced at a rate of $2.35\times10^9 \ \text{cm}^{-3} \ \mu \text{s}^{-1}$ resulting in a bulk plasma density of approximatey $n_e\approx5\times10^8 \text{cm}^{-3}$. The particle weights were $4\times10^9$, $1.6\times10^4$, and $2\times10^3$ for neutral helium, helium ions, and electrons respectively. The plasma was sourced with an ion temperature of 0.086 eV and electron temperature of 4 eV. The 2D electron temperature, which is the relevant temperature for comparison with the theory, is defined as $T_e= n_e\int d^3v m_e(v^2_{r,x}+v^2_{r,y})f_e/2$, where $v_{r,i}=(\vc{v}-\vc{V_e})\cdot \vc{\hat{i}}$.  The 2D electron and ion temperatures had a value of 1.64 eV and 0.048 eV near the sheath giving an electron sheath Bohm speed of 54.4 cm $\mu \text{s}^{-1}$. Only elastic collisions between ions and neutrals with a background pressure of 1 mTorr were included, and there was no volume generation of plasma. A $1\times10^{-4} \mu s$ time step was chosen to resolve the local electron plasma frequency throughout the domain. The simulation ran for $5\times10^6$ time steps resulting in $50\mu \text{s}$ of physical time. 

Two cases were considered, one where the electrode was biased +20 V with respect to the grounded walls, and the other with the electrode bias at -20 V. For the +20 V electrode an  electron sheath was allowed to form since the electrode satisfied $A_{E}/A_{W}<\sqrt{2.3 m_e/m_i}$ and the probe was biased above the plasma potential\cite{2007PhPl...14d2109B}. The potential gradient and electron and ion current vectors are shown in Fig.~6 for an electron sheath biased at +20 V and an ion sheath biased at -20 V. In the electron sheath case, the electron current indicates that the biased electrode has an effect on the bulk plasma that is significantly grater than that of the ion sheath. The importance of the 2D nature of the electron sheath can be seen in two effects. First, the current vectors of the repelled population (ions for the electron presheath and electrons for the ion presheath) are almost absent in the case of the ion presheath near the electrode, while those for the electron presheath indicate a significant flow velocity. For an infinite planar boundary it is not possible to have flow around the boundary. This is difficult to achieve for an electron sheath because the electrode must have a dimension that is small compared to the chamber size in order to be biased positive with respect to the plasma. The second effect is the convergence of the electron current into the electrode, even for distances greater than 1 cm away. This convergence is not seen for the ion presheath. The importance of these 2D effects will be explored in Sec. IIIC.

\newpage

\

\onecolumngrid

\

\
\begin{figure}[h]
\center
\includegraphics[scale=.48]{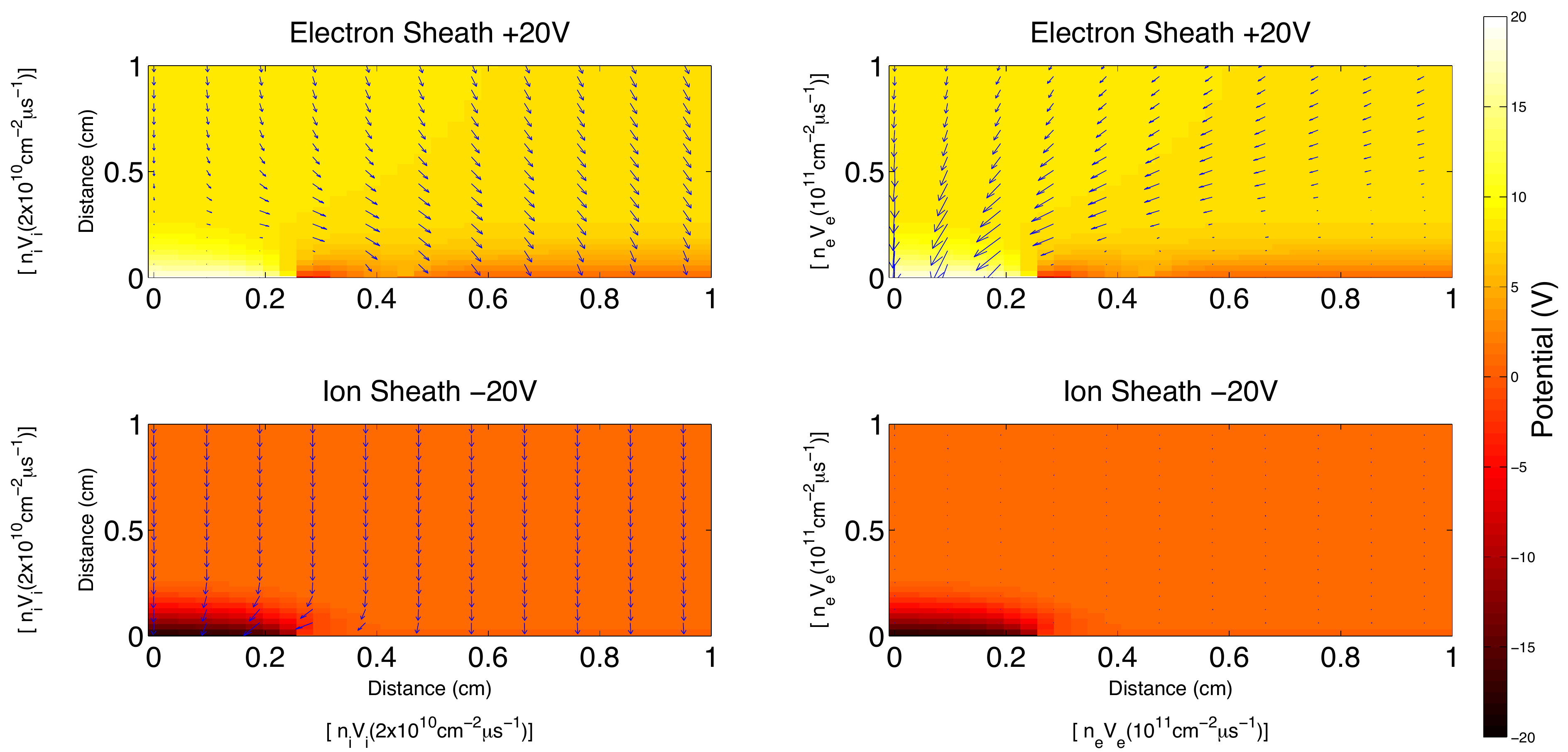}
\caption{The current flow vectors plotted on top of the potential for $n_iV_i$ (left column) and $n_eV_e$ (right column) for an electron sheath (top row) biased +20 V and ion sheath (bottom row) biased -20 V.  The electrode is between x = 0 and x = 0.25 cm, see annotations on Fig.~10 for more details of the simulation domain. All potentials are measured relative to the grounded wall. The greatest differences between these are that the electron presheath has a much greater effect on the electrons in the bulk plasma than the ion presheath has on the ions, and that the electron presheath redirects the ions, while the ion presheath has little effect on the electrons. Note the difference in scale for vectors in the left and right columns. }
\end{figure}
\twocolumngrid

\subsection{Electron fluid}

The simulations have shown that the electron sheath interfaces with the bulk plasma through a presheath. In this subsection simulations are compared to the presheath description given in the 1D model. Fig.~7 shows a comparison of the potential profile from the electron sheath PIC simulations and the models given for the presheath and sheath. For the presheath, the models for constant mean free path given in Eq.(8) and constant collision frequency in Eq.(10) were compared by fixing the value $\phi =0$ at the location where the electron sheath Bohm speed is attained, from here moving out some distance y into the plasma the potential profile was plotted. In the sheath, starting at the electrode, the sheath thickness as a function of potential from Eq.(16) was plotted for an argument $\Delta\phi=\phi_{E}-\phi(y)$ from $\phi(y)=\phi_{E}$ out to $\phi(y)=0$, the potential at which the electron sheath Bohm speed was attained. Here $\phi_E$ is the electrode potential. For comparison, the conventional model from Eq.(17) is also plotted. The potential profiles within the sheath are in excellent agreement with Eq.(16) indicating that the numerical factor corresponding to the flowing Maxwellian is the correct value. This result is significant since it indicates that the electron sheath is approximately twice as thick as was previously thought under the random flux assumption. The presheath potential profiles are plotted with a presheath length scale of $l = 0.3 \ \text{cm}$, which  approximately corresponds to the region in which Eq.(21) accurately describes the ion density in Fig.~7, as well as the region in Fig.~8a where the pressure gradient dominates over the electric field. The presheath potential profiles from the theory were shallower than that from the simulations near the sheath, however the slopes are in better agreement further away. This is possibly due to matching the simulation data at the theory's singular point. Simulation results only match the theory in a region where the electron presheath is dominant, however the model does not consider the interface of the presheath with a nonuniform bulk plasma such as the one in the simulations.

\begin{figure}[h!]
\centering
\includegraphics[scale=.45]{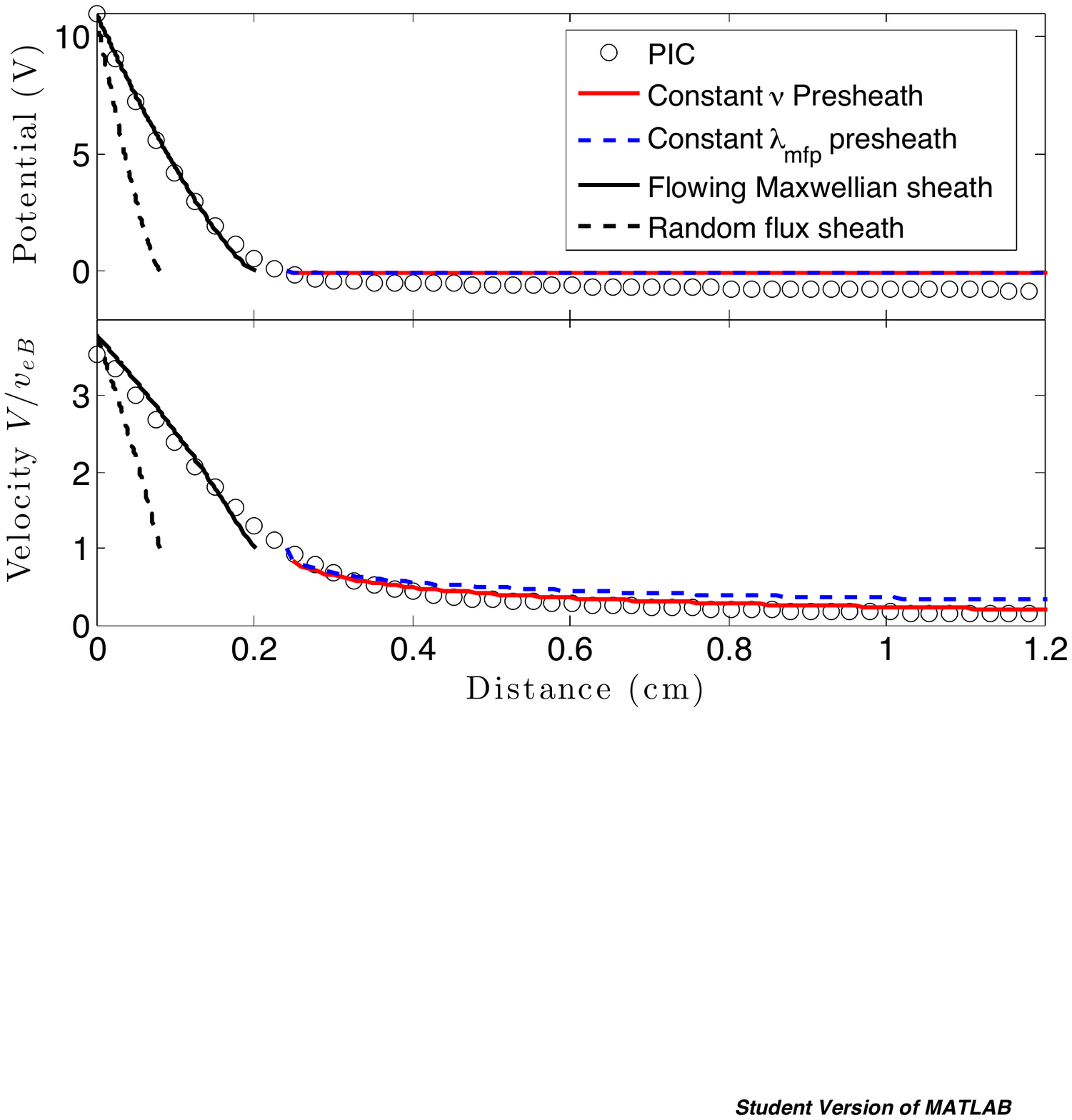}
\caption{Top: The potential profile from simulations compared to the model for the sheath and presheath. Bottom: Comparison of the flow velocity profile from simulation to the models for the sheath and presheath.  }
\end{figure}

The flow profiles of Eq.(7) and Eq.(9) are also compared to the simulations in Fig.~7. The flow profiles show that the sheath and presheath are in good agreement with theory. Due to presheath ion density fluctuations (see Fig.~14), the sheath edge is difficult to locate in the time averaged simulation data. To compare the electron flow velocity at the sheath edge two definitions are utilized, 1) the sheath thickness given by Eq. (16) for $\Delta\phi=\phi(0 \  \text{cm})-\phi(1 \ \text{cm})$, and 2) the location where the average difference between electrons and ions, $2(n_e-n_i)/(n_e+n_i)$ is greater than 30\%. The sheath edge is difficult to locate in the time averaged data. The chosen value of 30\% corresponds well with the typical sheath edge position in the time dependent data in Fig.~14 shown in Sec. IIID. By these two definitions the sheath edge is between 0.213 cm and 0.265 cm, the corresponding flow velocities are 1.21$v_{eB}$ and 0.85$v_{eB}$. Fig.~8a shows the ion and electron density, while Fig.~8b shows the corresponding terms in the electron momentum equation. Here the two dashed lines indicate the two sheath edge locations. In the region bounded by these two sheath edge definitions, the electric field overtakes the pressure gradient and the sheath begins. The location at which the electric field becomes the dominant driving term in the electron momentum equation closely coincides with the location at which the electron sheath Bohm velocity is achieved. 

Previously, in Sec. IIB, a comparison of presheath length scales was made for sheaths dominated by a common source of plasma generation between electrons and ions and no other collisions. For this situation it was concluded in Eq.(11) that the electron presheath was $\sqrt{m_i/m_e}$ longer than the ion presheath. In the simulations, no particles are sourced in the presheath so the dominant mechanism for determining the electron presheath length scale is expected to be electron-ion collisions. In the PIC simulations the electrons are collisionless in the Coulomb collision sense since electron-particle interactions are not considered within the computational cells. Another possible mechanism for such collisions are those due to particle wave interactions in an unstable plasma \cite{2010PhPl...17e5704B}. The collision rate due to electron interactions with ion acoustic waves has been important for explaining the anomalous scattering of electrons near the ion sheath\cite{2009PhRvL.103t5002B}, a phenomenon known as Langmuir's Paradox\cite{1955Natur.176..916G}. The EVDFs in Fig.~2 suggest a similar anomalous scattering mechanism may be important here since at the sheath edge and within the sheath electrons with velocities directed towards the bulk plasma are still present. Evidence for the presence of electron collisions can be obtained by adding up the terms in the momentum equation which are calculated from the simulations. In fact, one can see that the terms in Fig.~8b do not exactly cancel. Using PIC plasma quantities, the residual 
\begin{equation}
R_e= V_e\frac{d V_e}{d y}+\frac{e}{m_e}E+\frac{T_e}{m_e n_e}\frac{d n_e}{d y}
\end{equation}
was also plotted. An increase in the residual as the electrode is approached suggests that other neglected terms, (i.e. stress gradients, perpendicular velocity gradients, and friction) may be important. In particular a friction term may be due to wave particle interactions and could play an important role in determining the presheath length scale since it would determine the value of $\nu_R$ in Eq.(6). Instabilities will be discussed further in Sec. IIID.

\begin{figure}[h!]
\centering
\includegraphics[scale=.43]{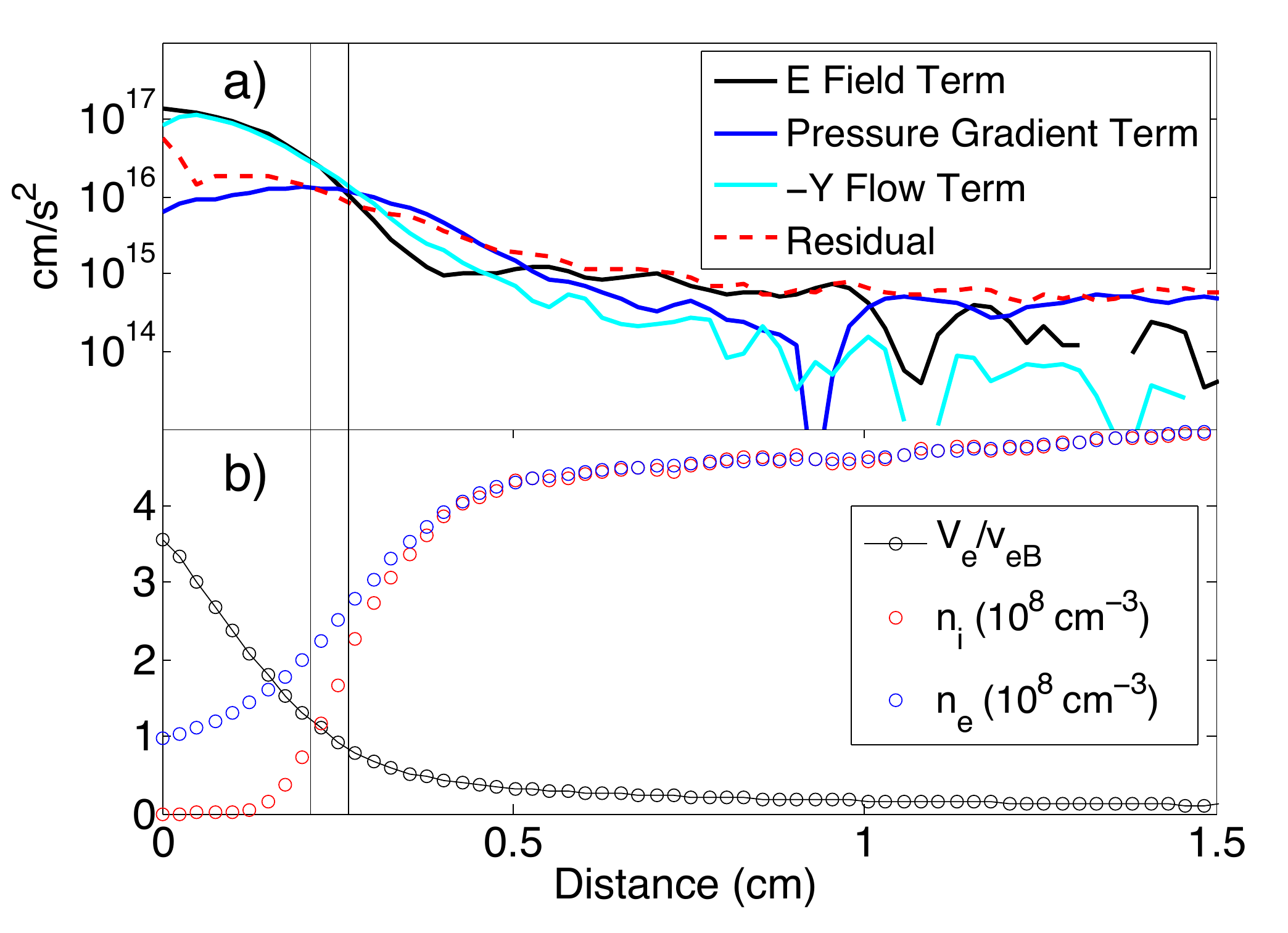}
\caption{a) Evaluation of the electron momentum equation terms using PIC simulation results. b) PIC simulation results for electron flow speed, ion density, and electron density. The two vertical lines indicate the sheath edge calculated as the location where $n_i$ and $n_e$ differ by 30$\%$ (left) and by the Child Langmuir law (right). }
\end{figure}

\subsection{Ions}

 \subsubsection{Ion density}
Plots of the ion current, in Fig.~6, show that ions flow around the electrode and are collected by the adjacent wall. Here the ion density will clearly be dominated by the flow profile around the electrode. This flow is a 2D effect that is absent in the description of ion sheaths near planar boundaries. Size limitations on the electron sheath from global current balance prevent it from being well described by an infinite 1D planar geometry. 
 
To model how the 2D flow affects the ion density profile, consider the 2D steady state ion momentum equation along a 1D cut perpendicular to the electrode center,
  \begin{equation}
m_in_i\bigg( V_x\frac{dV_y}{dx}+V_y\frac{dV_y}{dy}\bigg)=-en_i\frac{d\phi}{dy}-\frac{d}{dy}(n_iT_i).
\end{equation}
Here, the stress gradient and friction terms have been neglected. In Sec. IIB it was found that the electron presheath has weak potential gradients. Dropping the electric field term and integrating from the sheath edge back into the presheath results in 
 \begin{equation}
\frac{n_i(y)}{n_i(y_o)}=\exp\bigg[-\int_{y_o}^{y}\frac{m_i}{T_i}\bigg( V_x\frac{dV_y}{dx}+V_y\frac{dV_y}{dy}\bigg)dy\bigg],
\end{equation}
 where $y_o$ denotes the  sheath edge position. In this form the ion flow is balanced by the pressure gradient. This can be  contrast with the Boltzmann relation where the electric field and pressure gradient balance. The exact form of the pressure gradient is dependent on the electric field, after all it is the field that causes the density gradient. Determining the exact pressure gradient would involve solving the full 2D momentum equation with Poisson's equation using all the boundary conditions. For this section, numerical values from PIC simulations are used to test the relation in Eq. (21). Fig.~9 shows the presheath densities from PIC simulations compared to the evaluation of Eq. (21). These two quantities are in good agreement. For comparison the Boltzmann relation, with initial values in the presheath, is also shown in Fig.~9, demonstrating that Eq. (21) is a vast improvement in the description of the ion density. 
 \begin{figure}[h!]
\centering
\includegraphics[scale=.4]{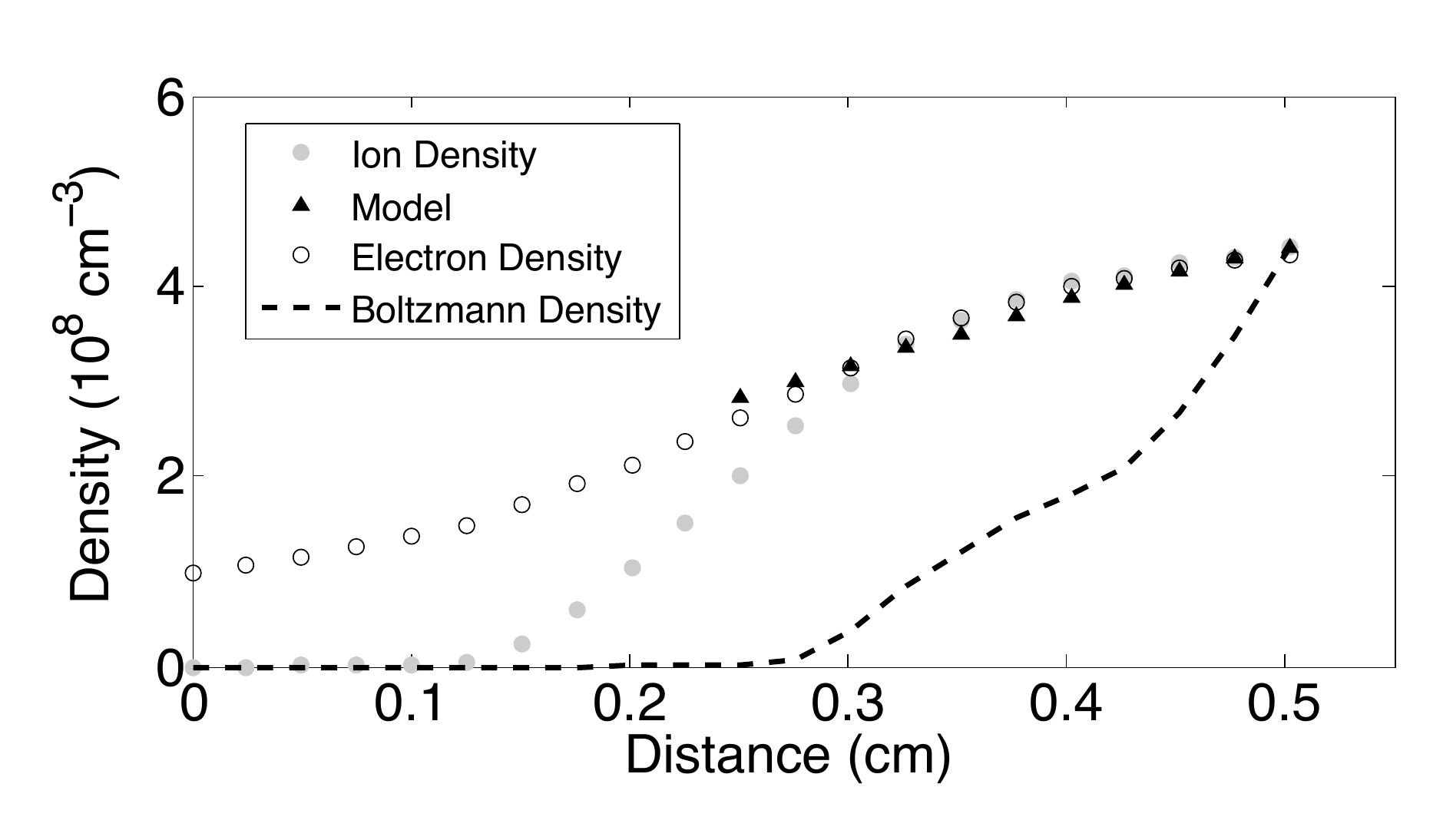}
\caption{The electron and ion density for a 1D cut in the simulation domain perpendicular to the electrode. The ion density integral of Eq. (21) evaluated for PIC velocity and temperature profiles (blue triangles) agrees well. }
\end{figure}
 \subsubsection{Ion VDFs} 
 The effect of the electron sheath on ions can also be explored from a kinetic point of view. Fig.~10 shows ion heating in the electron presheath. This heating is explained as a result of ion interaction with the presheath. This interaction generates a flow moment in the ion VDFs (IVDFs) in the transverse direction when approaching the electrode. The 2D IVDFs, are shown in Fig.~11. These demonstrate that the majority of ions are redirected away from the boundary and collected by the adjacent grounded wall. It is this redirection that is primarily responsible for the heating, however there is also a small population of ions that are reflected back into the plasma. 
 \begin{figure}[h!]
\centering
\includegraphics[scale=.38]{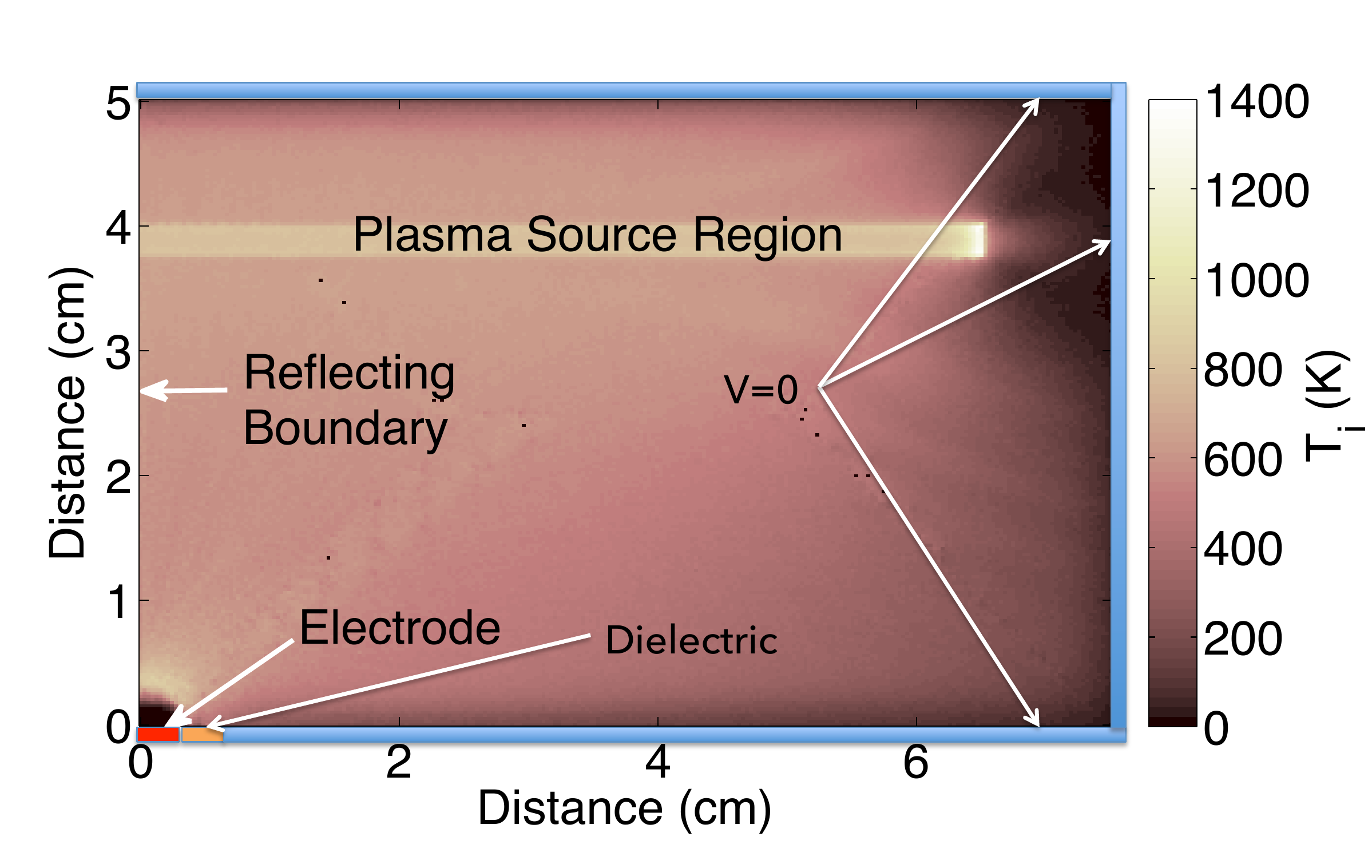}
\caption{The ion temperature throughout the simulation domain. The plasma source region, electrode, reflecting boundary, grounded walls, and dielectric are marked. Note the heating in the region just above the electron sheath in the lower left corner of the domain.}
\end{figure}
 
The 2D IVDFs in Fig.~11 were computed in the presheath using individual particle positions and velocities over 30 $\mu \text{s}$ and were averaged over 0.1 cm $\times$ 0.1 cm boxes starting at the sheath edge around 0.25 cm moving back into the plasma 0.85 cm. The averaging boxes also extend the length of the electrode in the x direction, with the last box including the electrode wall boundary. Far from the electrode the IVDFs are flow shifted towards the boundary, as would be expected for an expanding plasma, and show little modification apart from a small population of reflected ions. As the ions approach the electrode some of their flow velocity is diverted from the -y to x direction since the ions are repelled by the 2D presheath electric field which has x and y components. 
 \begin{figure}[h!]
\centering
\includegraphics[scale=.7]{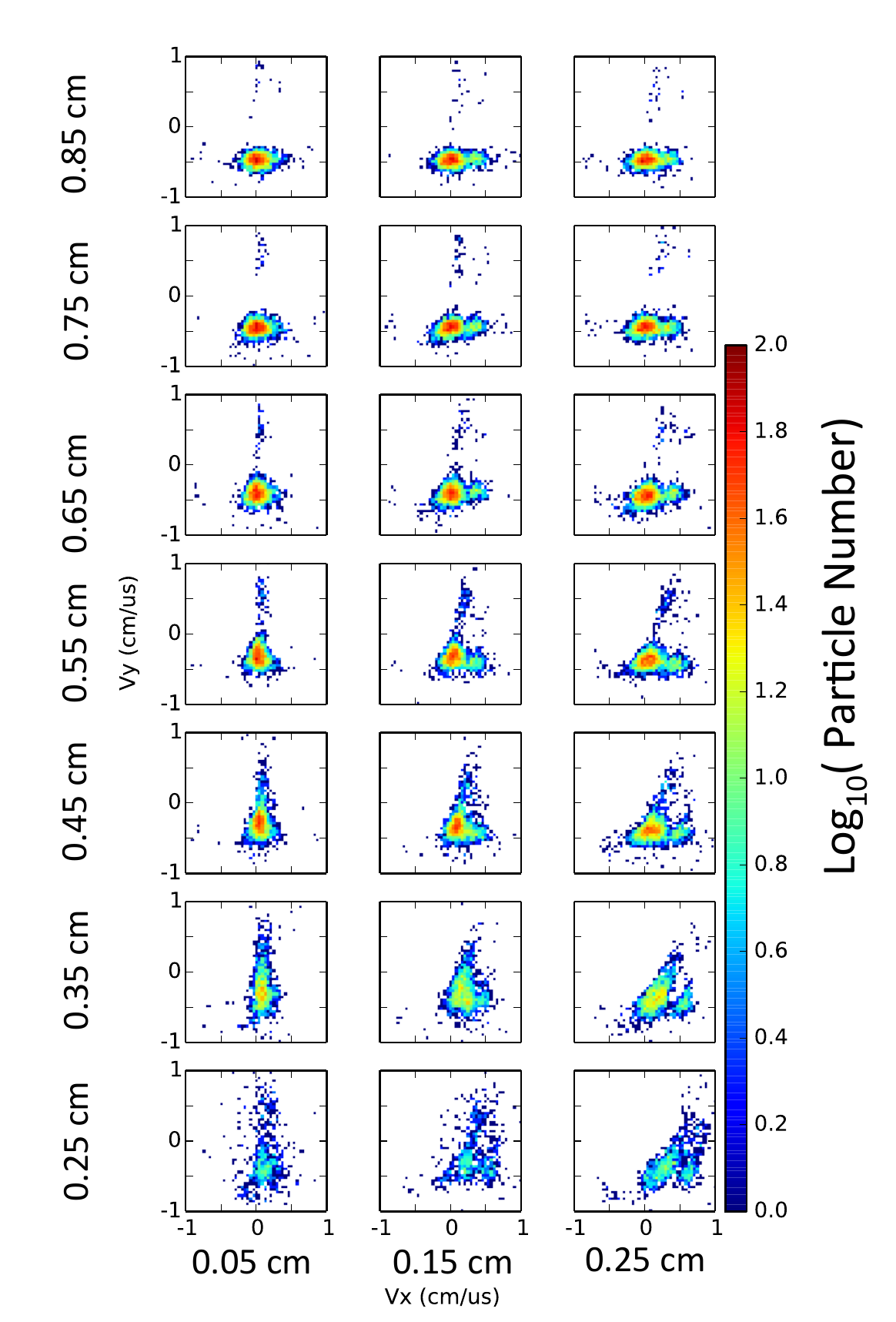}
\caption{Ion VDFs near the electron sheath biased +20V above ground shown in Fig.~6. The IVDFs were averaged over 0.1 cm x 0.1cm boxes. The labels in the x and y axes indicate the coordinate of the center of the box, the electrode is on the x axis at y=0 between x=0 and 0.25 cm (averaging starts 0.2 cm above the electrode), since further below there are not enough ions for meaningful IVDFs.  }
\end{figure}

The modification of the IVDF shape near the electrode can be described by the flow around the electrode. Consider the IVDFs halfway between the plasma source and the boundary containing the electrode, each starting at three different locations in the x direction, see the location marked A in Fig.~12(a). At the starting location each IVDF will have a flow due to the plasma expansion, so the distribution will have a flow shift in the direction of the electrode or wall, this is represented in Fig.~12(b). Due to the flow around the electrode each of these distributions will end at the location marked B in Fig.~12(a). Now consider the distribution with flow incident on the electrode. Since the flow is redirected the flow shift of this distribution will be transferred from the -y direction to the x direction as it approaches the electrode. Likewise, a distribution incident to the edge of the anode will also have its flow diverted from the -y direction to the x direction, although to a lesser extent. Finally, a distribution incident to the grounded wall will remain unchanged. The final position of these three distributions is shown in Fig.~12(c), although a more realistic expectation would be smeared out, such as the distribution shown in Fig.~12(d), due to a continuum of starting positions.

\begin{figure}[h]
\centering
\includegraphics[scale=.35]{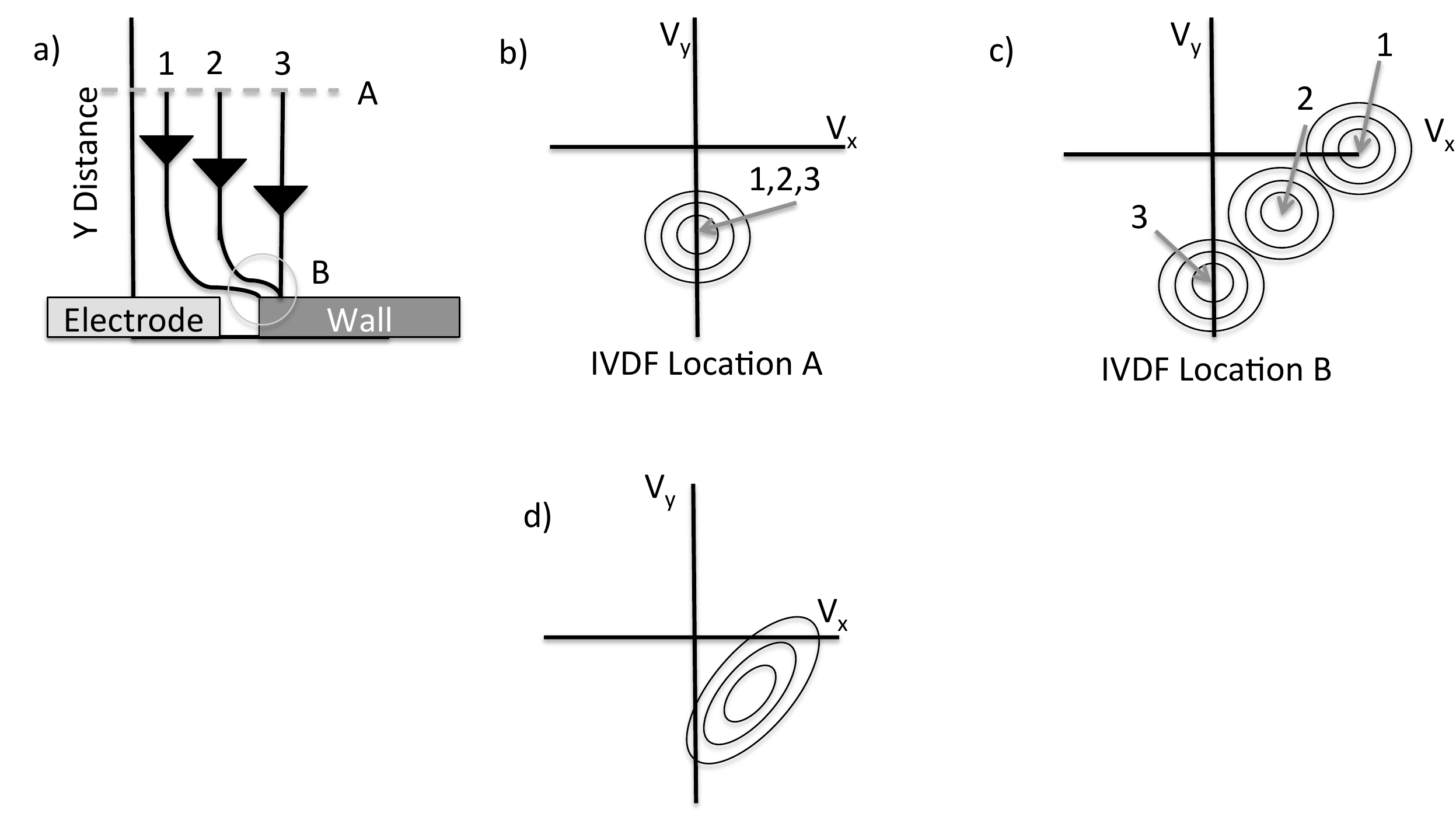}
\caption{Schematic drawing describing the time-averaged IVDFs at different locations in the plasma. a) The flow lines of particles in 3 different VDFs starting at A and ending at B. b) The VDFs at location A. c) The VDFs at location B, distributions incident on the electrode have their -y velocity redirected in the x direction. d) A realistic IVDF at location B due to a continuum of starting positions along line A.}
\end{figure}

The basic expectations of the model shown in Fig.~12(d) are borne out in the simulated IVDFs near the boundary in Fig.~11. It is important to note that the physical picture illustrated in Fig.~12 is not exact since not every particle flows along a stream line, but experiences diffusion as well. There are small scale features not explained by the picture in Fig.~12. For instance, in some IVDFs there is a small secondary maximum to the right of the primary. This situation may be due to time averaging of the particle positions and velocities over 30 $\mu s$ in combination with fluctuations in the presheath caused by instabilities. 

\subsection{Fluctuations and instabilities}

The simulated electron sheath, shown in the 1 cm$\times$1 cm panels of Fig.~13, exhibits fluctuations of the sheath edge position on the order of 0.05 cm on a time scale of approximately 1 $\mu s$. Fluctuations were not observed for the ion sheath with the electrode biased at -20 V. The presence of a differential flow, approaching the electron thermal speed, between electrons and ions in the electron presheath is expected to give rise to ion-acoustic instabilities for the present values of $T_i/T_e$. In this subsection the effect of these waves on the fluctuations is explored. Two-dimensional FFTs of the ion density confirm that the sheath fluctuations are due to ion acoustic waves.

\begin{figure}[h!]
\centering
\includegraphics[scale=.4]{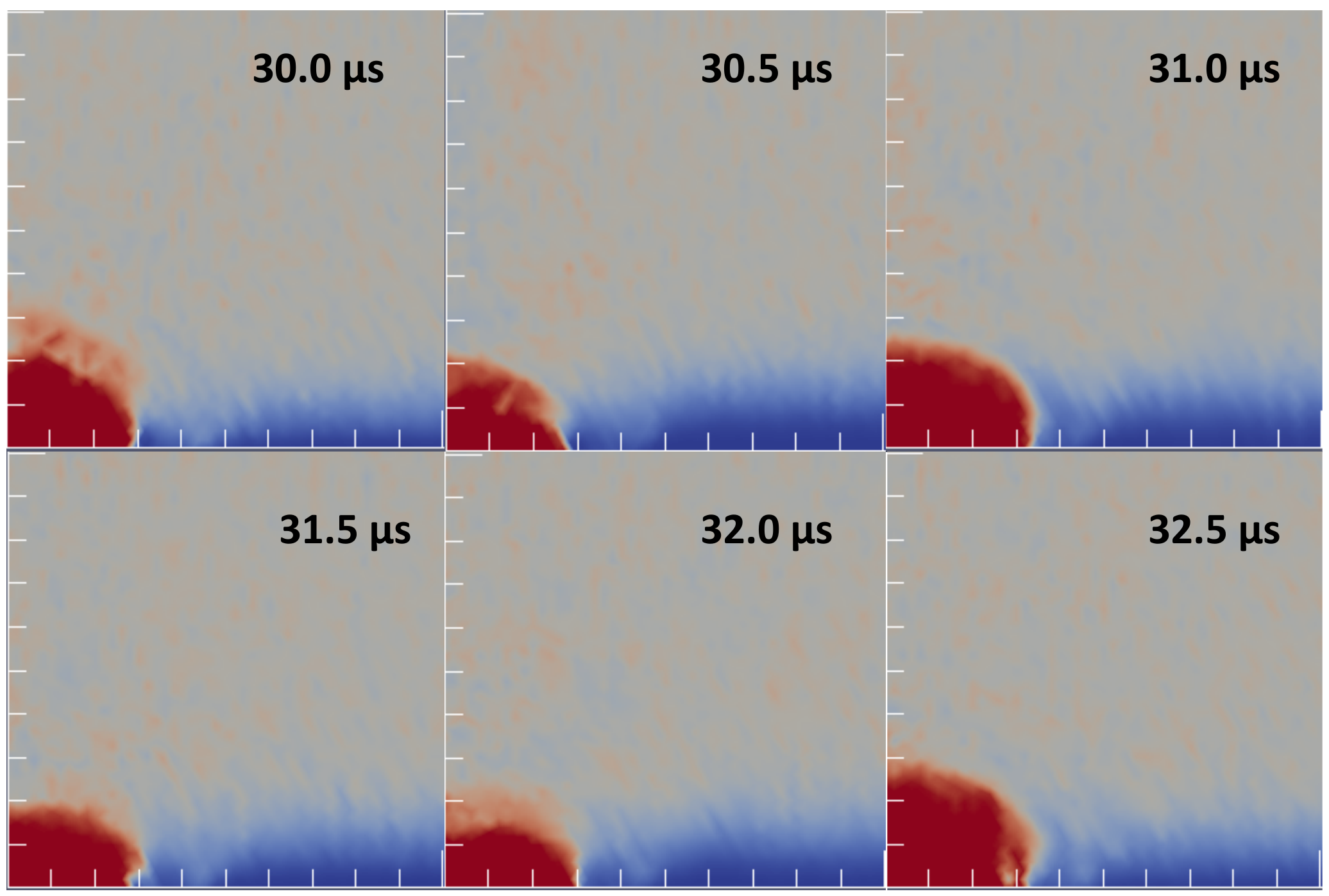}
\caption{The fluctuation of the 2D electron sheath boundary plotted in a 1 $\times$ 1 cm region at $0.5 \mu s$ intervals.The color indicates charge density, with red being electron rich, and blue being ion rich.}
\end{figure}

The dielectric response for a plasma where the electrons are Maxwellian with flow $V_e$ and stationary Maxwellian ions is\cite{1984bpp..conf..519D}
\begin{equation}
\vc{\epsilon}(\vc{k},\omega)=1-\frac{\omega_{pe}^2}{k^2 v_{T_e}^2}Z^\prime(\xi_e)-\frac{\omega_{pi}^2}{k^2 v_{T_i}^2}Z^\prime(\xi_i)
\end{equation}
where $\xi_e=\frac{\omega -\vc{k}\cdot\vc{V_e}}{kv_{T_e}}$ and $\xi_i=\frac{\omega}{kv_{T_i}}$, and $Z^\prime$ is the derivative of the plasma dispersion function\cite{1961pdf..book.....F}. The dispersion relation is determined by the zeros of the dielectric function. The approximate solution is 
\begin{equation}
\frac{\omega}{\omega_{pi }}\approx \frac{k\lambda_{D_e}}{\sqrt{k^2\lambda^2_{D_e}-\frac{1}{2}Z^\prime\big(-\frac{V_e}{v_{T_e}}\big)}},
\end{equation}
which was determined by expanding the ion term using the large argument expansion of the plasma dispersion function\cite{1983nrl..reptQ....B} and by using the approximations $\omega/k\sim c_s$ and $V_e \gg c_s$ in the electron term.

 Fig.~14 shows the ion density along a line extending 1 \ cm perpendicular to the electrode over a $5 \ \mu s$ interval. The figure shows that there are ion density fluctuations that propagate towards the sheath edge as time increases. The figure also shows that the sheath edge position fluctuations closely follow the propagation of the ion density fluctuations, meaning that these are likely responsible for the sheath edge and resulting current fluctuations which are associated with positively biased probes\cite{1981PlPh...23..325G,1978JPSJ...44..991D}. The 2D FFT of the ion density shown in Fig.~15 was computed over a line extending 1 cm from the electrode. These FFTs were examined to determine whether or not the ion density fluctuations are ion acoustic waves. The FFTs are in fair agreement with the expected dispersion relation determined from  Eq. (23) indicating that the density disturbances, which are responsible for the sheath edge fluctuations, are in fact ion acoustic waves. The figure also indicates that nonlinear effects may be producing a cascade to shorter scales.

\begin{figure}[h!]
\centering
\includegraphics[scale=.58]{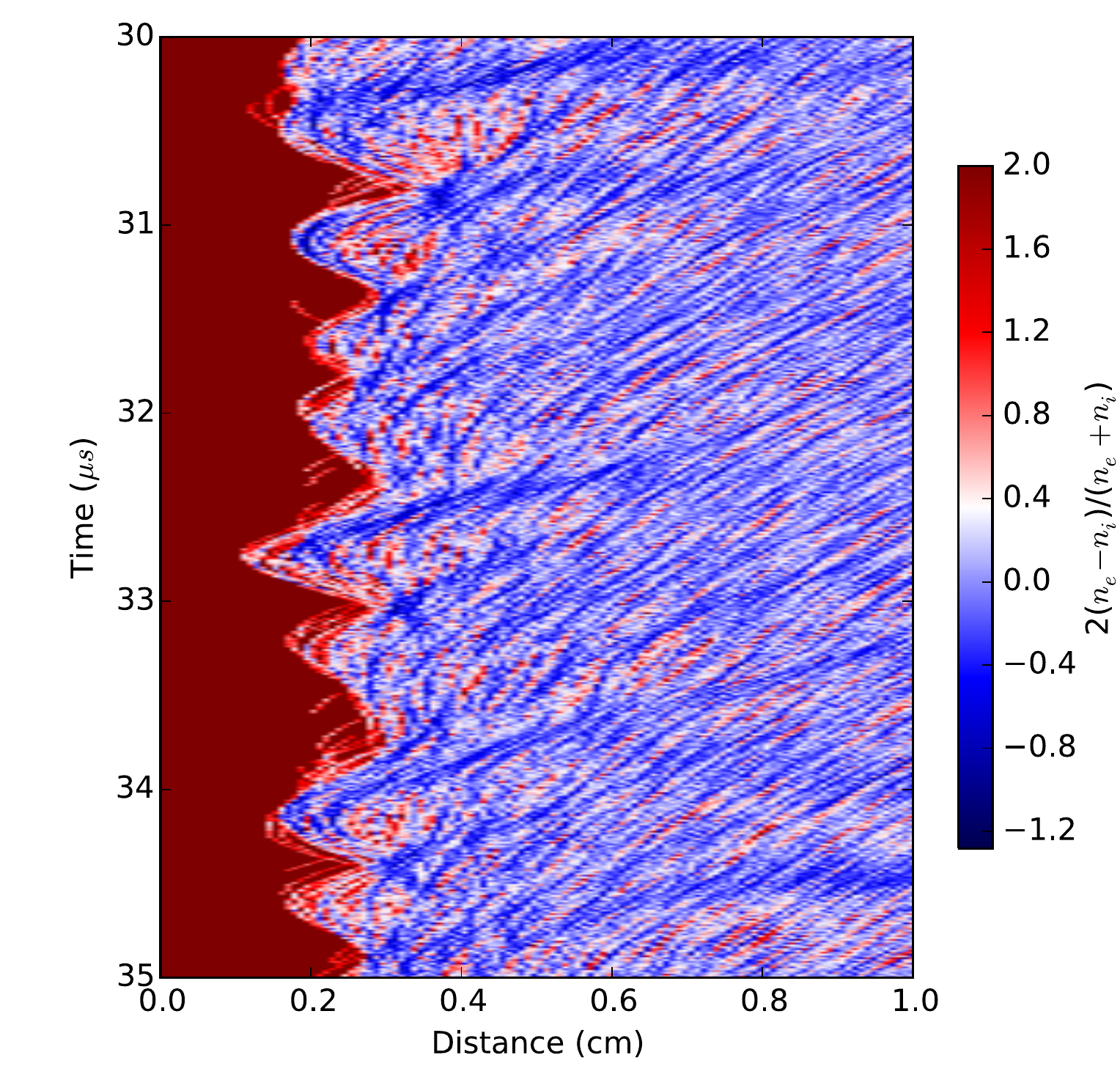}

\includegraphics[scale=.62]{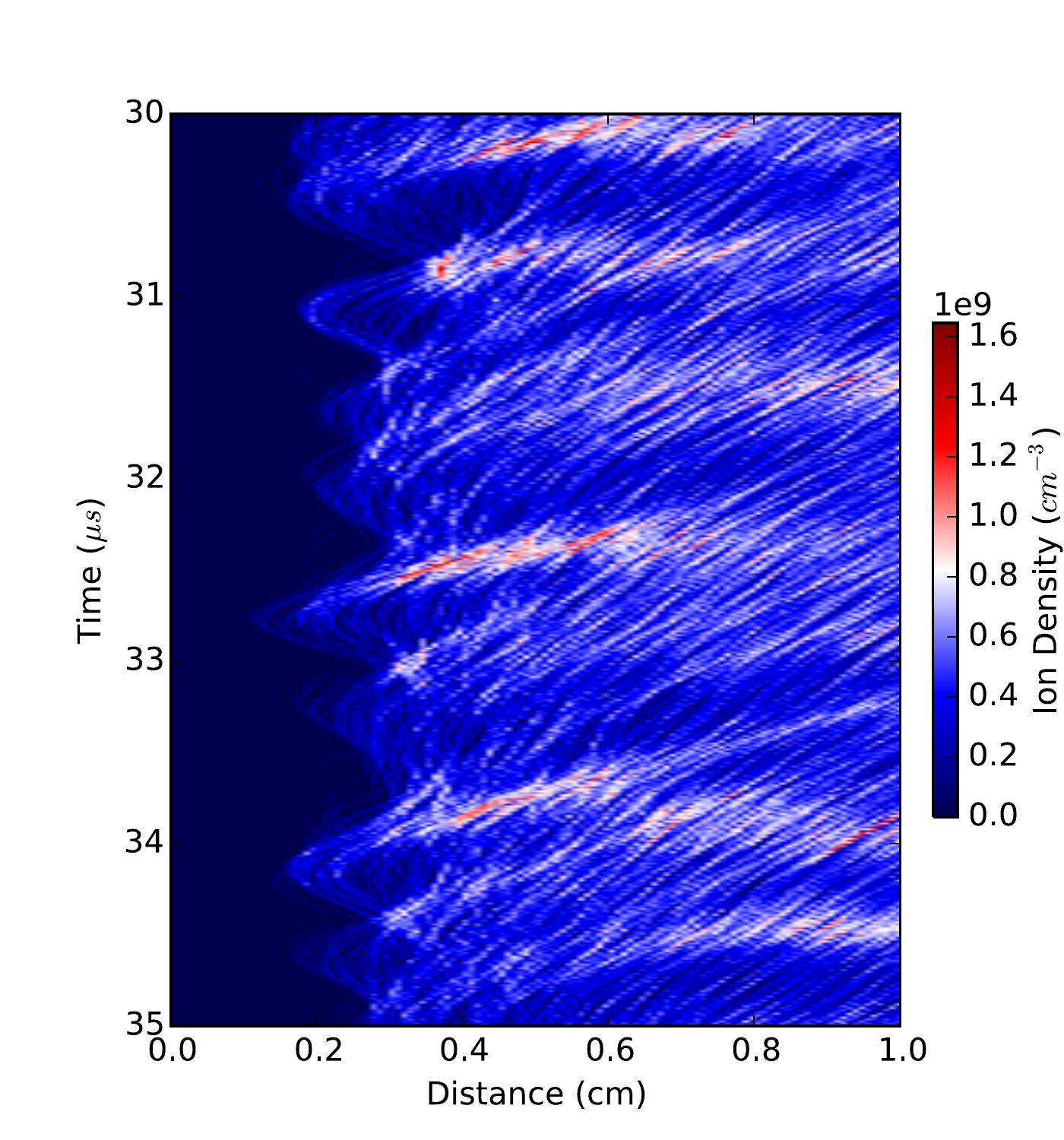}
\caption{
Top: The absolute difference between the electron and ion density is plotted to show the sheath edge position fluctuations, measured along an axis perpendicular to the electrode, as a function of time over a 5$\mu s$ interval. Bottom: The ion density over the same time interval. The ion density fluctuations correspond to the sheath edge fluctuations. }
\end{figure}

\begin{figure}[h!]
\centering
\includegraphics[scale=.38]{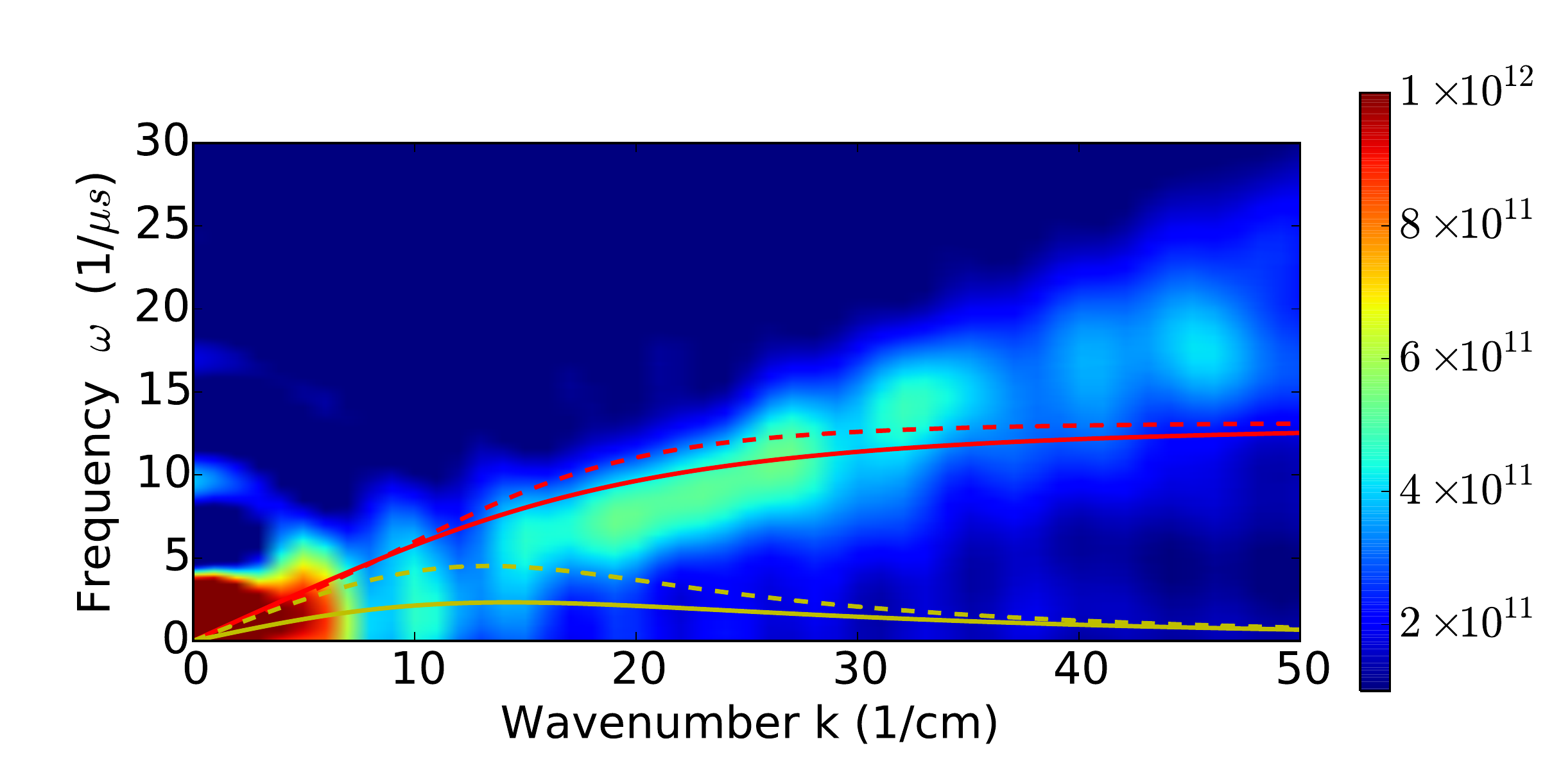}
\caption{ The 2D FFT of the ion density shown in Fig.~14. The solid and dashed red lines corresponds to the real part of the approximate dispersion relation given in Eq. (23) for electron flows of $0.5v_{eB}$ and $0.9 \ v_{eB}$. The yellow and dashed yellow lined show the imaginary part of Eq. (23). }
\end{figure}

\section{SUMMARY}
In this paper the conventional picture that the electron sheath collects a random flux of electrons was shown to be incomplete. Based on the EVDFs of 2D PIC simulations, a model was developed using the electron momentum and continuity equations where the EVDF is a flowing Maxwellian. In this model the electron sheath interacts with the bulk plasma through a presheath where the electron velocity approaches the electron sheath Bohm speed, $\sqrt{(T_e+T_i)/m_e}$. In this presheath there are shallow potential gradients that drive a large pressure gradient. It is this pressure gradient that is primarily responsible for the acceleration of electrons. 

The 1D model was compared to the 2D simulations using the time averaged values from the simulation. Within the sheath the potential profiles and flow velocities are in excellent agreement with the flowing Maxwellian model, which results in an electron sheath that is approximately twice as thick as the one described by the commonly assumed random flux model. The simulations are consistent with the electron flow velocity attaining the electron sheath Bohm speed by the sheath edge, and this flow velocity was verified to be the result of acceleration in a pressure driven electron presheath. Comparison with the simulations also revealed the inherent 2D nature of the electron sheath. Due to it's small size, the electron presheath does not resemble the presheath of an infinite planar boundary, instead there is a divergence of the ion flow around the sheath-presheath region. This flow necessitates a new description of ions where the ion flow is balanced by the presheath pressure gradients. 

Finally, the simulations revealed the existence of ion density fluctuations in the electron presheath. These density fluctuations are expected; The theory predicts a large differential flow between ions and electrons in the presheath which excite ion acoustic instabilities. FFTs of the 2D ion density indicate that these density fluctuations are ion acoustic waves. Inspection of the sheath edge position revealed that these ion acoustic waves are responsible for sheath edge fluctuations, and hence sheath collection area fluctuations, which in turn cause fluctuations in the collected electron saturation current.



\section*{Acknowledgments}
This research was supported by the Office of Fusion Energy Science at the U.S. Department of Energy under contract
DE-AC04-94SL85000. The first author was also supported by the U.S. Department of Energy, Office of Science, Office of Workforce Development for Teachers and Scientists, Office of Science Graduate Student Research (SCGSR) program. The SCGSR program is administered by the Oak Ridge Institute for Science and Education for the DOE under contract number DE-AC05-06OR23100.

\section*{Appendix A: Exact Solutions to Eq. (15)}
The solution to an equation of the form 
\begin{equation}
y^2-2\ln(y)=z
\end{equation}
can be written in terms of the Lambert W function as 
\begin{equation}
y=\exp\bigg\{-\frac{1}{2}W\bigg(-\frac{1}{e^z}\bigg)-\frac{z}{2}\bigg\}. 
\end{equation}
The Lambert W function has two branches, the $W_0(z)$ branch and the $W_{-1}(z)$ branch. For the electron sheath problem we are interested in the asymptotic limit as $z\to\infty$. For this limit the $W_0(z)$ branch provides unphysical solutions because $W_0(0)=0$  and an accelerating flow velocity cannot correspond to $y\to0$, instead we choose the $W_{-1}$ branch. The  asymptotic limit of the $W_{-1}(z)$ branch as $z\to 0^-$ is\cite{Corless96onthe}
\begin{eqnarray}
W_{-1}(z)=\ln(-z)-\ln(-\ln(-z)) \nonumber \\
+\mathcal{O}\bigg(\frac{\ln(-\ln(-z))}{\ln(-z)}\bigg).
\end{eqnarray}
Using the asymptotic limit in the solution Eq. (24) gives $y=\sqrt{z}$, the same result as if the logarithmic term were dropped. To quantify the error involved in this approximation we plot Eq. (24) against $\sqrt{z}$ in Fig.~16, and see that the error is $\approx 20\%$ at small z and decreases at large z.
\begin{figure}[h!]
\centering
\includegraphics[scale=0.43]{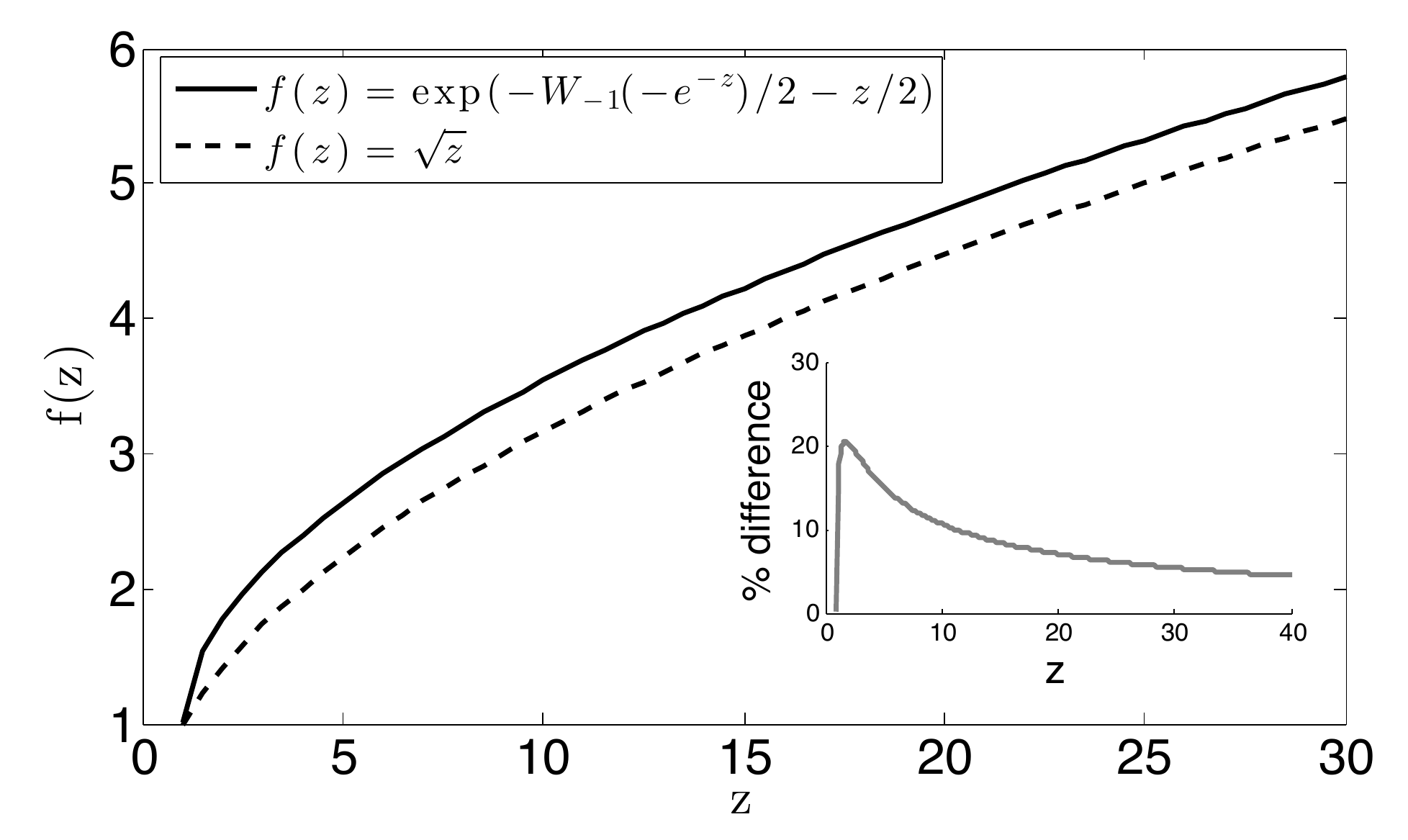}
\caption{Exact solution to (24) and the solution using the asymptotic approximation of $W_{-1}(z)$ as $z\to 0^-$. The subplot shows the percent difference of the two curves. }
\end{figure}

\bibliography{draft9}

\end{document}